\documentclass[11pt,a4paper]{article} 

\usepackage{simplewick}

\usepackage[showcomments]{ajtex}

\usepackage{tikz}
\usetikzlibrary{shapes.misc}

\newcommand{\ftikz}[2][0]{\raisebox{#1pt}{\includegraphics[]{tikz/#2.pdf}}}

\def\bx{{\bf x}}
\def\bk{{\bf k}}
\def\bq{{\bf q}}

\def\GGL{\text{GGL}}

\usepackage{tikz}
\usetikzlibrary{decorations.markings,decorations.pathmorphing,arrows.meta}

\usetikzlibrary{external}
\tikzset{aux/.style={decorate,decoration={snake,segment length=1.5mm,amplitude=0.4mm}}}
\tikzset{bold/.style={ultra thick}}
\tikzset{cdotted/.style={line width=1.2pt, dash pattern=on 0pt off 1.2mm, line cap=round}}
\tikzset{bcdotted/.style={line width=2pt, dash pattern=on 0pt off 1.5mm, line cap=round}}

\tikzset{left/.style={arrows={Stealth[scale length=0.5, scale width=1.5][sep=2pt]-}}}
\tikzset{right/.style={arrows={-[sep=-2pt]Stealth[scale length=0.5, scale width=1.5]}}}
\tikzset{lleft/.style={arrows={Stealth[scale length=0.5, scale width=1.5][sep=2pt]-}}}
\tikzset{right/.style={arrows={-[sep=-2pt]Stealth[scale length=0.5, scale width=1.5]}}}

\newcommand\farc[2]{
\draw [#1] (0,0)--(-3.5mm,3.5mm) arc (225:85:4.5mm);
\draw [#2] (0,0)--(3.5mm,3.5mm) arc (-45:90:4.5mm);
\draw [right] (-0.1mm,11mm)--(0,11mm);
}
\newcommand\floop[4]{
\draw [#1] (0,0) arc (180:90:5mm);
\draw [#2] (10mm,0) arc (0:90:5mm);
\draw [#3] (0,0) arc (180:270:5mm);
\draw [#4] (10mm,0) arc (0:-90:5mm);
\draw [right] (5mm,5mm)--(5.1mm,5mm);
\draw [right] (5mm,-5mm)--(4.9mm,-5mm);
}

\newcommand*\fline[4][5mm]{
\draw [#2] (#4,0)--(\dimexpr#4+#1\relax,0);
\draw [#3] (\dimexpr#4+#1\relax,0)--(\dimexpr#4+#1+#1\relax,0);
\draw [right] (\dimexpr#4+#1-1mm\relax,0)--(\dimexpr#4+#1\relax,0);
}
\newcommand\fextin[2]{
\IfStrEq{#1}{2}
    {\draw [>>-] (\dimexpr#2-2.5mm\relax,0)--(#2,0);}
    {\draw [>-] (\dimexpr#2-1.5mm\relax,0)--(#2,0);}
}
\newcommand\fextout[2]{
\IfStrEq{#1}{2}
    {\draw [->>] (#2,0)--(\dimexpr#2+2.5mm\relax,0);}
    {\draw [->] (#2,0)--(\dimexpr#2+2mm\relax,0);}
}
\newcommand*\flabel[2][1]{
\IfStrEq{#1}{0}{
\node at (0mm,7mm) {\textcolor{darkgray}{\footnotesize $#2$}};
}{
\node at (5mm,0) {\textcolor{darkgray}{\footnotesize $#2$}};
}
}

\newcommand\retdiagramA[9][]{
~\raisebox{-15pt}{
\tikz[thick]{
    \fextin{}{-10mm};
    \fline{#2}{#3}{-10mm};
    \floop{#4}{#5}{#7}{#6};
    \fline{#8}{#9}{10mm};
    \fextout{2}{20mm};
    \flabel{#1};
}}~
}

\newcommand\retdiagramAp[7][]{
~\raisebox{-15pt}{
\tikz[thick]{
    \fextin{}{-10mm};
    \fline{#2}{#3}{-10mm};
    \floop{#4}{#5}{#7}{#6};
    \draw (10mm,0) -- (11mm,0);
    \fextout{2}{11mm};
}}~
}

\newcommand\retdiagramApp[7][]{
~\raisebox{-15pt}{
\tikz[thick]{
    \fextin{}{0mm};
    \floop{#2}{#3}{#5}{#4};
    \fline{#6}{#7}{10mm};
    \fextout{2}{20mm};
}}~
}

\newcommand\retdiagramAppp[5][]{
~\raisebox{-15pt}{
\tikz[thick]{
    \fextin{}{0mm};
    \floop{#2}{#3}{#5}{#4};
    \draw (10mm,0) -- (11mm,0);
    \fextout{2}{11mm};
}}~
}

\newcommand\retdiagramC[7][]{
~\raisebox{-1pt}{\tikz[thick]{
    \fextin{}{-10mm};
    \fline{#2}{#3}{-10mm};
    \farc{#4}{#5};
    \fline{#6}{#7}{0mm};
    \fextout{2}{10mm};
}}~
}

\newcommand\retdiagramCp[5][]{
~\raisebox{-1pt}{\tikz[thick]{
    \fextin{}{-10mm};
    \fline{#2}{#3}{-10mm};
    \farc{#4}{#5};
    \draw (0,0) -- (1mm,0);
    \fextout{2}{1mm};
}}~
}

\newcommand\retdiagramCpp[5][]{
~\raisebox{-1pt}{\tikz[thick]{
    \fextin{}{-.5mm};
    \draw (-.5mm,0) -- (0,0);
    \farc{#2}{#3};
    \fline{#4}{#5}{0mm};
    \fextout{2}{10mm};
}}~
}

\newcommand\retdiagramCppp[3][]{
~\raisebox{-1pt}{\tikz[thick]{
    \fextin{}{-.5mm};
    \draw (-.5mm,0) -- (0,0);
    \farc{#2}{#3};
    \draw (0,0) -- (1mm,0);
    \fextout{2}{1mm};
}}~
}

\newcommand\aline[2]{

\FPeval\xval{cos(#2*\FPpi/2)}
\FPeval\yval{sin(#2*\FPpi/2)}

\IfStrEq{#1}{ext}{
    \draw [>-] (2*\xval mm,2*\yval mm)--(0,0);
}{
    \IfStrEq{#1}{ext2}{
        \draw [>>-] (3.2*\xval mm,3.2*\yval mm)--(0,0);
    }{
        \draw [#1] (0,0)--(8*\xval mm,8*\yval mm);
        \draw[left] (8.5*\xval mm,8.5*\yval mm)--(8*\xval mm,8*\yval mm);
    }
}
}

\newcommand\twoint[2]{
\hspace{0.5em}
\raisebox{-1pt}{\tikz[thick]{
    \aline{#1}{0};
    \aline{#2}{2};
}}
\hspace{0.5em}
}

\newcommand\threeint[4][fill]{
\raisebox{-19pt}{\tikz[thick]{
    \node at (6mm,6mm) {\textcolor{white}{$.$}};
    \node at (-6mm,-6mm) {\textcolor{white}{$.$}};
    \aline{#2}{0.666};
    \aline{#3}{2};
    \aline{#4}{3.333};
}}
}

\newcommand\fourint[5][fill]{
\raisebox{-19pt}{\tikz[thick]{
    \node at (6mm,6mm) {\textcolor{white}{$.$}};
    \node at (-6mm,-6mm) {\textcolor{white}{$.$}};
    \aline{#2}{0.5};
    \aline{#3}{1.5};
    \aline{#4}{2.5};
    \aline{#5}{3.5};
}}
}
  
\title{Quantum fluctuations in hydrodynamics and quantum long-time tails}

\author{Akash Jain}\email{akash.jain@maths.ox.ac.uk}

\affiliation{Mathematical Institute, University of Oxford, Woodstock Road, Oxford, OX2 6GG, U.K.}

\abstract{
We construct a quantum Schwinger-Keldysh (SK) effective field theory for the diffusive hydrodynamics of a conserved scalar field. Quantum corrections within the SK framework are guided by fluctuation-dissipation relations, enforced via a dynamical Kubo-Martin-Schwinger (KMS) symmetry. We find that the KMS symmetry necessarily generates fluctuation contributions in the SK effective action at all orders in the noise field, thereby giving rise to intrinsically non-Gaussian noise. We use our results to compute one-loop quantum corrections to the two-point density-density retarded correlation function, leading to a quantum generalization of hydrodynamic long-time tails. Our results apply at arbitrarily high orders in $\hbar$. The one-loop results for retarded correlation functions have been expressed in terms of a family of polynomials. We also provide a closed-form expression for the one-loop results at leading order in the wavevector expansion.
}

\setlist[itemize]{itemsep=0pt,topsep=0pt}

\usepackage{xfp}

\usetikzlibrary{external}
\begin{document} 

\hypersetup{linktoc=all}
\setcounter{tocdepth}{3}
\maketitle

\section{Introduction}

Fluctuation-dissipation theorem (FDT) requires that a physical system with dissipation must also exhibit fluctuations. Hydrodynamics describes the long-wavelength dissipative dynamics of many-body systems close to thermal equilibrium. However, the traditional framework of hydrodynamics---built on conservation laws and deterministic constitutive relations---does not account for fluctuations mandated by FDT~\cite{landau1959fluid}. Fluctuations at finite temperature persist even in the classical limit, arising as stochastic uncertainty due to coarse-graining over microscopic degrees of freedom. Stochastic hydrodynamics accounts for these classical thermal fluctuations by incorporating random noise into the constitutive relations~\cite{1963AnPhy..24..419K, Martin:1973zz, Pomeau:1974hg, Hohenberg:1977ym, DeDominicis:1977fw, Khalatnikov:1983ak}. However it still fails to account for the full quantum nature of fluctuations as systems depart from the classical limit.

Hydrodynamics is organised as a perturbative expansion near thermal equilibrium. Given the frequency $\omega$ of fluctuations near equilibrium, hydrodynamics requires that $\omega \ll 1/t_{\text{micro}}$, where $t_{\text{micro}}$ is largest characteristic timescale of the microscopic theory, like the mean free time between collisions of constituent particles. On the other hand, quantum fluctuations operate at the time scale $\hbar\beta$, where $\kB\beta=1/T$ is the inverse temperature of the system. For theories where there is a sensible separation of scales between between $t_{\text{micro}}$ and $\hbar\beta$, we may systematically ignore the quantum fluctuations. This is the regime of validity of stochastic hydrodynamics.
However for systems where all intrinsic timescales are either comparable to or smaller than $\hbar\beta$, there is no meaningful way to separate quantum and stochastic fluctuations.
An important example are conformal field theories (CFTs), which lack any characteristic scale of their own and thus $t_{\text{micro}}$ is set precisely by the thermal scale $\hbar\beta$.\footnote{This applies to CFTs with finite number of colours $N$. For large-$N$ CFTs, the microscopic time-scale is instead controlled by $t_{\text{micro}}\sim N\hbar\beta \gg \hbar\beta$. I thank Shiraz Minwalla for pointing this out.}

Schwinger-Keldysh (SK) hydrodynamics is an action-based effective field theory (EFT) framework for hydrodynamics~\cite{Grozdanov:2013dba, Harder:2015nxa, Crossley:2015evo, Haehl:2015uoc, Haehl:2018lcu, Jensen:2017kzi}, which enables us to systematically include the effects of thermal as well as quantum fluctuations. In particular, the framework features a symmetry-based realisation of the FDT, known as the dynamical Kubo-Martin-Schwinger (KMS) symmetry. This symmetry acts as a guiding principle for introducing fluctuations in hydrodynamics, both thermal and quantum, consistent with the FDT requirements at the full non-linear level.

The action of the KMS symmetry on SK-EFTs is non-local in time. SK-EFTs are set up on a closed-time contour that goes back-and-forth in time~\cite{Bellac:2011kqa}, with independent degrees of freedom and background fields living on both parts of the contour. We will collectively denote these as $f_1(t)$ and $f_2(t)$. The KMS symmetry acts on these fields with a relative phase-shift $i\hbar\beta$ in imaginary time, i.e.
\begin{align}
    f_1(t) \to \pm f_1(-t+i\theta), \qquad 
    f_2(t) \to \pm f_2(-t - i\hbar\beta +i\theta).
\end{align}
where $\theta$ is a constant parameter and $\pm$ denotes the time-reversal eigenvalue of the fields. As a result, any coupling terms in the effective action between $f_1$ and $f_2$ transform non-locally in time under KMS symmetry. This makes the KMS symmetry quite hard to implement in SK-EFTs. 

Interestingly, the KMS symmetry becomes local in the classical limit $\hbar\to0$. To this end, we identify the average ``physical'' combination of the fields $f_r = (f_1+f_2)/2$ and the difference ``noise'' combination $f_a = (f_1-f_2)/\hbar$. These transform under the KMS symmetry at leading order in $\hbar$ as
\begin{align}
    f_r(t)\to\pm f_r(-t) + \cO(\hbar), \qquad 
    f_a(t)\to \pm f_a(-t) \pm i\beta\dow_tf_r(-t)
    + \cO(\hbar).
\end{align}
This classical version of the KMS symmetry is relatively easier to implement in SK-EFTs and has been successfully applied to derive an EFT for stochastic hydrodynamics. 

In this work, we present a systematic mechanism to implement the full quantum version of the KMS symmetry in the SK effective action. This construction enables us to derive the quantum-complete SK-EFT for hydrodynamic models, consistent with FDT requirements at finite $\hbar$. We focus on a simple hydrodynamic model of a single conserved scalar field for concreteness, though extensions to more general hydrodynamic models is straight-forward. As an application of our framework, we derive one-loop quantum corrections to the two-point retarded correlation funtion of density. This enables us to compute the quantum correction to hydrodynamic long-time tails in the theory of diffusion.

This paper is organised as follows. In the rest of this introduction, we will provide an overview of the main technical results of this work. In \cref{sec:SK-formalism}, we give a brief review of the SK-EFT formalism for diffusion, the KMS symmetry, and the SK effective action in the classical limit. \Cref{sec:quantum-action} forms the main body of this paper, where we construct the quantum complete version of the SK effective action consistent with the KMS symmetry. We use these results in \cref{sec:retarded-correlators} to obtain one-loop quantum corrections to the two-point retarded correlation function of density. We close with some discussion in \cref{sec:outlook}. Some of the details of Feynman rules in the presence of sources have been relegated to \cref{app:feynman-rules-source}.

\subsection{Overview of results}

\paragraph*{The classical diffusion model:}

For concreteness, we will work with the simplest hydrodynamic model for a single conserved density $n$ with diffusive dynamics. This simple model can describe conserved particle dynamics, conserved U(1) charges, or conserved energy in the absense of momentum conservation. The extension to full hydrodynamic models with conserved momenta is conceptually straightforward.
The classical conservation equation for $n$ is given as
\begin{align}\label{eq:conservation}
    \dow_t n + \dow_i J^i = 0,
\end{align}
where $J^i$ denotes the flux. Instead of working directly with $n$, it is convenient to instead express the theory in terms of the associated thermodynamic chemical potential $\mu$, such that $n(\mu)$ is a function of $\mu$. The flux $J^i$ admits diffusive constitutive relations that take the form
\begin{align}\label{eq:constitutive}
    J^i = -\sigma(\mu)\lb \dow^i\mu - E^i \rb,
\end{align}
where $\sigma(\mu)$ is the conductivity transport coefficient and $E_i=\dow_iA_t-\dow_tA_i$ denotes an applied electric field. Note that the response of $J^i$ to applied electric fields as well as chemical potential gradients is controlled by the same transport coefficient $\sigma$. This follows as a consequence of the second law of thermodynamics; see e.g.~\cite{Jain:2023obu} for a quick derivation. 

Substituting the constitutive relations into the conservation equation and linearising in fluctuations $\delta n = n - n(\mu_0)$, we see that the model features diffusive dynamics $\dow_t \delta n = D\,\dow_i\dow^i\delta n - \sigma\, \dow_i E^i$, where $D=\sigma/\chi$ is the diffusion constant. In Fourier-space parametrised by frequency $\omega$ and wavevector $k_i$, this leads to the standard diffusion pole $\omega = -iD\bk^2$. We can also use the hydrodynamic model to obtain the classical retarded correlation functions (or response function). For instance, the two-point classical retarded correlation function of density is given as
\begin{align}
    G^\rmR_{nn}(\omega,\bk)_{\text{cl}} 
    = \frac{\delta n_{\text{onshell}}[A]}{\delta A_t}
    \bigg|_{A=0}
    = \frac{\sigma\,\bk^2}{D\,\bk^2 - i\omega},
\end{align}
which admits a diffusive pole. See~\cite{Kovtun:2012rj} for more details.

\paragraph*{Classical SK-EFT for diffusion:}

The degrees of freedom in the SK-EFT for diffusion are a pair of phase fields $\varphi_1$ and $\varphi_2$. The average field $\varphi_r = (\varphi_1+\varphi_2)/2$ is related to the chemical potential via $\mu = \dow_t\varphi_r + A_t$. Whereas, the difference combination $\varphi_a = (\varphi_1-\varphi_2)/\hbar$ is understood as the noise field parametrising fluctuations. The SK effective action for diffusion takes the form
\begin{align}
    S = \int \df t\,\df^d\bx 
    \Big( n(\mu) \dow_t\varphi_a 
    - \sigma(\mu)\lb \dow^i\mu - E^i \rb \dow_i\varphi_a
    + \cO(\varphi_a^2)
    \Big),
    \label{eq:action-simple}
\end{align}
where we have suppressed the quadratic and higher-order terms in the noise field. Here $d$ denotes the number of spatial dimensions.
Extremising the effective action with respect to $\varphi_r$ or $\mu$ sets the noise field $\varphi_a$ to zero onshell. Whereas, extremising it with respect to $\varphi_a$ leads the the classical conservation equation in \cref{eq:conservation} with the constitutive relations in \cref{eq:constitutive}.

To progress beyond the classical equations of motion and account for fluctuations, we need to specify the higher-order noise terms in the effective action \eqref{eq:action-simple}. At the classical level, the KMS symmetry determines these to be~\cite{Crossley:2015evo, Chen-Lin:2018kfl, Jain:2020zhu}
\begin{align}
    S = \int \df t\,\df^d\bx 
    \lb n(\mu) \dow_t\varphi_a 
    - \sigma(\mu)\lb \dow^i\mu - E^i \rb \dow_i\varphi_a
    + \frac{i}{\beta}\sigma(\mu)\, \dow_i\varphi_a\dow^i\varphi_a
    + \cO(\hbar)
    \rb.
    \label{eq:action-cl-simple}
\end{align}
Note that the noise part of the effective action is imaginary, which concerns the dissipative nature of SK-EFT models. We can use this model to obtain the retarded as well as symmetric correlation functions in the classical limit. To this end, we need to also couple the effective action to noise-type gauge fields $A_{at}$, $A_{ai}$, which is achieved by making the substitutions $\dow_t\varphi_a\to\dow_t\varphi_a + A_{at}$ and $\dow_i\varphi_a\to\dow_i\varphi_a + A_{ai}$ in \cref{eq:action-cl-simple}. For uniformity, in the SK-EFT language it is customary to denote the original gauge fields $A_{t}$, $A_i$ as $A_{rt}$, $A_{ri}$. 

At tree-level, i.e. ignoring interactions in the SK-EFT, \cref{eq:action-cl-simple} leads to two-point retarded and symmetric correlation functions of density
\begin{alignat}{2}
    G^\rmR_{nn}(\omega,\bk)
    &= \bfrac{\delta}{\delta A_{rt}}
    \bfrac{-i\delta}{\delta A_{at}}
    \ln{\cal Z}
    \bigg|_{A_r,A_a=0}
    &&= \frac{\bk^2\sigma}{D\,\bk^2 - i\omega}
    + \cO(\hbar)
    + \ldots
    , \nn\\
    G^\rmS_{nn} (\omega,\bk)
    &= \bfrac{-i\delta}{\delta A_{at}}
    \bfrac{-i\delta}{\delta A_{at}}
    \ln{\cal Z}
    \bigg|_{A_r,A_a=0}
    &&= \frac{2}{\beta}
    \frac{\bk^2\sigma}{|D\,\bk^2 - i\omega|^2}
    + \cO(\hbar)
    + \ldots,
    \label{eq:corr-from-gf-intro}
\end{alignat}
where ellipses denote loop-corrections due to interactions. For the retarded correlation function, we have recovered the classical result in \cref{eq:corr-from-gf} at treee level. Whereas, flucutations in the model also lead to a non-trivial symmetric correlation function, which parametrises the probabilistic variance of the density, $\langle n^2\rangle - \langle n\rangle^2$. Note that the symmetric correlation function has poles at $\omega = \pm iD\bk^2$. The correlation functions satisfy the classical FDT~\cite{Callen:1951vq}
\begin{equation}
    G^\rmS_{nn}(\omega,\bk)
    = \frac{2}{\beta\omega}
    \Im G^\rmR_{nn}(\omega,\bk)
    + \cO(\hbar),
    \label{eq:fdt-cl-n}
\end{equation}
which is ensured by the KMS symmetry of \cref{eq:action-cl-simple}.

We can also use \cref{eq:action-cl-simple} to calculate the corrections to the correlation functions due to interactions with classical fluctuations. For example, the leading correction to the two-point retarded correlation function of density is given as~\cite{Chen-Lin:2018kfl,Jain:2020zhu}
\begin{align}
    G^\rmR_{nn}(\omega,\bk)
    =
    \frac{\bk^2\sigma}{D\,\bk^2 - i\omega} 
    + \frac{i\omega\,\bk^4}{(D\,\bk^2 - i\omega)^2}
    \frac{\chi^2\lambda^2}{\beta D} 
    \frac{\Gamma\bigl(1-\tfrac d2\bigr)}{(16\pi)^{d/2}}
    \lb \bk^2 - \frac{2i\omega}{D}\rb^{\tfrac{d}{2}-1}
    + \cO(\hbar)
    + \ldots.
    \label{eq:long-time-tails}
\end{align}
In particular, note that the correction piece is non-analytic and gives rise to a branch-cut in the complex-$\omega$ plane starting at $\omega = -iD\bk^2/2$. In real-space, this leads to a power-law fall-off of the correlation function at late times, known as long-time tails~\cite{Forster:1976zz, Kovtun:2003vj}, instead of the exponential decay as predicted by classical deterministic hydrodynamics. Note that in $d=2$, the non-analyticity behaves as $\ln \lb \bk^2 -2i\omega/D\rb$ instead.

\paragraph*{Quantum-complete SK-EFT for diffusion:}

The main result of this paper is to derive the quantum-completion of the SK effective action in \cref{eq:action-cl-simple}, guided by the finite-$\hbar$ version of the KMS symmetry. We find that the KMS symmetry generates terms in the effective action at all orders in the noise field. Truncating to quartic order in the noise field, we find
\begin{align}\label{eq:quantum-action-simple}
    S
    &= \int \df t\,\df^d\bx \bigg[
    n(\mu) \dow_t\varphi_a
    - \sigma(\mu)\lb \dow^i\mu - E^i \rb \dow_i\varphi_a \nn\\
    &\qquad
    + \frac{i}{\beta} \sigma(\mu) \dow_i\varphi_a \fD \dow^i\varphi_a
    - \frac{i}{\beta} \lb \dow^i\mu - E^i \rb\,
    \fD_2\Big(\dow_i\varphi_a,\sigma'(\mu) \dow_t\varphi_a \Big)
    \nn\\
    &\qquad
    + \frac{\hbar^2}{24} n''(\mu)(\dow_t\varphi_a)^3
    - \frac{\hbar^2}{8} \sigma''(\mu) \lb \dow^i\mu - E^i \rb \dow_i\varphi_a (\dow_t\varphi_a)^2
    - \frac{\hbar^2}{8} 
    \dow_t\Big(\sigma'(\mu) \dow_t\varphi_a \Big) 
    \dow_i\varphi_a \dow^i\varphi_a
    \nn\\
    &\qquad
    + \frac{i\hbar^2}{8\beta} \sigma''(\mu)
    (\dow_t\varphi_a)^2 \dow_i\varphi_a \fD \dow^i\varphi_a
    - \frac{i\hbar^2}{24\beta} 
    \lb \dow^i\mu - E^i \rb
    \fD_2\Big(\dow_i\varphi_a,\sigma'''(\mu)  (\dow_t\varphi_a)^3\Big) \nn\\
    &\qquad
    + \cO(\varphi_a^5) \bigg],
\end{align}
where $\cD$ and $\cD_2$ are differential operators defined as
\begin{align}
    \fD = \frac{\hbar\beta\dow_t}{2}
    \cot\bfrac{\hbar\beta\dow_t}{2} 
    &= 1 - 2 \sum_{n=1}^{\infty} \zeta(2n)
    \bfrac{\hbar\beta\dow_t}{2\pi}^{2n},
    \nn\\
    &= 1 - \frac{\hbar^2\beta^2}{12}\dow_t^2
    - \frac{\hbar^4\beta^4}{720}\dow_t^4
    + \ldots, \nn\\[0.5em]
    \fD_2(f,g)
    = \frac{1}{\dow_t}\Big( g\fD f - f\fD g\Big) 
    &= 2\sum_{m=1}^{\infty}
    \sum_{n=1}^{2m} 
    (-1)^{n+1}\,
    \zeta(2m)
    \bfrac{\hbar\beta}{2\pi}^{2m}
    \dow_t^{n-1} f\,\partial_t^{2m-n}g \nn\\
    &= 
    \frac{\hbar^2\beta^2}{12}
    \lb  f\,\partial_t g - \dow_t f\,g \rb \nn\\
    &\qquad
    + \frac{\hbar^4\beta^4}{720}
    \lb f\dow_t^3 g - \dow_t f\dow_t^2 g + \dow_t^2 f \dow_t g 
    - \dow_t^3 g
    \rb 
    + \ldots,
\end{align}
where $\zeta$ denotes the Riemann Zeta function. Note that both these operators are perfectly local in time within a derivative expansion. Note that the quantum corrections leave the linear piece in $\varphi_a$ untouched. Therefore they do not affect the classical conservation equation of diffusion. However, they do affect the quadratic and all higher-order terms in $\varphi_a$. We will provide a detailed derivation of the quantum-complete SK effective action in \cref{sec:quantum-action}, along with a non-linear closed form valid for arbitrarily high powers of $\varphi_a$.

The SK effective action \eqref{eq:quantum-action-simple} leaves the retarded correlation function untouched at the tree-level. However, the symmetric correlation function modifies as
\begin{alignat}{2}
    G^\rmR_{nn}(\omega,\bk)
    &= \bfrac{\delta}{\delta A_{rt}}
    \bfrac{-i\delta}{\delta A_{at}}
    \ln{\cal Z}
    \bigg|_{A_r,A_a=0}
    &&= \frac{\bk^2\sigma}{D\,\bk^2 - i\omega}
    + \ldots
    , \nn\\
    G^\rmS_{nn} (\omega,\bk)
    &= \bfrac{-i\delta}{\delta A_{at}}
    \bfrac{-i\delta}{\delta A_{at}}
    \ln{\cal Z}
    \bigg|_{A_r,A_a=0}
    &&=
    \hbar\omega
    \coth\bfrac{\hbar\beta\omega}{2} 
    \frac{\bk^2\sigma}{|D\,\bk^2 - i\omega|^2}
    + \ldots,
    \label{eq:corr-from-gf-intro-quantum}
\end{alignat}
As a consequence of the KMS symmetry, these expressions satisfy the quantum FDT
\begin{equation}
    G^\rmS_{nn}(\omega,\bk)
    = \hbar
    \coth\bfrac{\hbar\beta\omega}{2}
    \Im G^\rmR_{nn}(\omega,\bk).
    \label{eq:fdt-q-n}
\end{equation}
In addition to diffusive poles at $\omega=\pm iD\bk^2$, the quantum-version of the symmetric correlation function also has Matsubara poles at $\omega = \pm im\pi/(\hbar\beta)$ for $m\in\bbZ^+$.

\paragraph*{Quantum long-time tails:}

As an application of the quantum-complete SK-EFT in \cref{eq:quantum-action-simple}, we can compute the one-loop quantum corrections to the two-point retarded correlation function of density in \cref{eq:long-time-tails}. We find
\begin{align}
    G^\rmR_{nn}(\omega,\bk)
    &=
    \frac{\bk^2\sigma}{D\,\bk^2 - i\omega} 
    + \frac{i\omega\,\bk^4}{(D\,\bk^2 - i\omega)^2}
    \frac{\chi^2\lambda^2}{\beta D} 
    \frac{\Gamma\bigl(1-\tfrac d2\bigr)}{(16\pi)^{d/2}}
    \lb \bk^2 - \frac{2i\omega}{D}\rb^{\tfrac{d}{2}-1}
    \nn\\
    &\hspace{8em}
    \times
    \lb 1
    - 2\sum_{n=1}^{\infty}
    \zeta(2n)
    \bfrac{\hbar\beta}{8\pi}^{2n} 
    \lb D\bk^2-2i\omega\rb^{2n}
    \cP_{d,2n}(z)
    \rb
    + \ldots,
    \label{eq:long-time-tails-quantum}
\end{align}
where $\lambda = \dow D/\dow n$ and $\cP_{d,2n}(z)$ are certain $2n$-degree polynomials of $z=\bk^2(\bk^2 - 2i\omega/D)^{-1}$, given in \cref{sec:one-loop}. At the time of writing, we are unable to identify any standard polynomials in the literature that match $\cP_{d,2n}(z)$. We observe that the quantum correction only show up at even powers of $\hbar\beta$.

Note that the individual polynomials do not give any additional non-analytic structure to the correlation function. However, the entire resummed expression does give rise to new Matsubara-like pole structures. While we have not been able to perform the resummation for arbitrary values of $\omega$ and $\bk^2$, we can obtain an exact answer for $D\bk^2 \ll |\omega|$, We note that the polynomials $\cP_{d,2n}(z)$ are normalised such that $\cP_{d,2n}(0)=1$, which leads to the closed form
\begin{align}
    \lb 1
    - 2\sum_{n=1}^{\infty}
    \zeta(2n)
    \bfrac{\hbar\beta}{8\pi}^{2n} 
    \lb D\bk^2-2i\omega\rb^{2n}
    \cP_{d,2n}(z)
    \rb
    = \frac{\hbar\beta \omega}{4}
    \coth\bfrac{\hbar\beta \omega}{4} 
    + \cO\big(D\bk^2/|\omega|\big).
\end{align}
Interestingly, note that the argument of coth in this expression is $\hbar\beta\omega/4$ unlike $\hbar\beta\omega/2$ in the symmetric correlation function in \cref{eq:corr-from-gf-intro-quantum}. This leads to additional Matsubara-like poles at $\omega = \pm 4im\pi/(\hbar\beta)$ for $m\in\bbZ^+$. We suspect that the poles in the upper-half complex-$\omega$ plane are removed upon including finite $\bk^2$ corrections. We leave these explorations for future work.

% \begin{align}
%     G^\rmR_{nn}(t,\bk)
%     &=
%     \int \frac{\df\omega}{2\pi}\E{-i\omega t} 
%     \frac{\bk^2\sigma}{D\,\bk^2 - i\omega} \nn\\
%     &\qquad 
%     + \frac{\chi^2\lambda^2}{4 D} 
%     \frac{\Gamma\bigl(1-\tfrac d2\bigr)}{(16\pi)^{d/2}}
%     \int \frac{\df\omega}{2\pi}\E{-i\omega t} 
%     \frac{i\hbar\omega^2\,\bk^4}{(D\,\bk^2 - i\omega)^2}
%     \lb \frac{-2i\omega}{D}\rb^{\tfrac{d}{2}-1}
%     \coth\bfrac{\hbar\beta \omega}{4}
%     + \ldots,
% \end{align}

% $\cP_{(2n)} = -\cB^{(2n-1)}(p)$.

\paragraph*{Covariant notation:}

We will use the covariant notation for coordinates $(x^\mu)\equiv (t,x^i)$ and momenta $(p_\mu) \equiv  (\omega,\bk)$ in the rest of this work. Similarly, we can define the covariant current $J^\mu \equiv (n,J^i)$ and background gauge field $A_\mu \equiv (A_t,A_i)$. However, we emphasise that this is purely a notation and does not imply that the theory is relativistic.

\section{Lightening review of Schwinger-Keldysh hydrodynamics}
\label{sec:SK-formalism}

We begin with a brief review of the Schwinger–Keldysh effective field theory (SK-EFT) formalism~\cite{Crossley:2015evo, Haehl:2018lcu, Jensen:2017kzi}, largely following the comprehensive treatment in~\cite{Glorioso:2018wxw}. In \cref{sec:SK-formalism-general}, we introduce the essential ingredients of the SK-EFT for diffusion, including the dynamical field content, background fields, and symmetries. The dynamical KMS symmetry is discussed in \cref{sec:KMS}. Finally, in \cref{sec:classical-action}, we explain how these elements are assembled to construct the SK effective action for diffusion in the $\hbar\to0$ limit. This analysis sets the stage for our construction of the SK effective action at finite $\hbar$ in \cref{sec:quantum-action}.

\subsection{Schwinger-Keldysh formalism}
\label{sec:SK-formalism-general}

\paragraph*{Dynamical fields and global symmetries:} 

SK-EFTs are set-up on a closed-time contour, with leg ``1'' going forward in time and leg ``2'' returning backward in time to the initial state. While the conventional single-time contour from textbook quantum field theory can only access correlation functions at zero temperature, this exotic closed-time contour can be used to compute symmetric, retarded, and advanced correlators at finite temperature.
Each leg of the contour is equipped with its own set of degrees of freedom and background fields. For the theory of diffusion, the dynamical field content consists of a pair of phase fields $\varphi_{1,2}$, with the subscripts labelling the respective legs of the contour.
The theory is required to obey independent global U(1) symmetries on the two legs, i.e.
\begin{subequations}
\begin{equation}
    \varphi_{1,2}\to \varphi_{1,2} - \Lambda_{1,2}.
\end{equation}
It is useful to introduce a set of background gauge fields $A_{1,2\mu}$ associated with these symmetries, with transformation rules
\begin{equation}
    A_{1,2\mu} \to A_{1,2\mu} + \dow_\mu \Lambda_{1,2}.
\end{equation}
\label{eq:global-U1-SK}%
\end{subequations}
Using these, we can define the global symmetry invariants
\begin{equation}
    B_{1,2\mu} = \dow_\mu\varphi_{1,2} + A_{1,2\mu},
\end{equation}
which are invariant under the background-gauged versions of the global U(1) symmetries.

It is also useful to introduce an average-difference basis 
\begin{align}
    f_r = \frac{f_1+f_2}{2}, \qquad 
    f_a = \frac{f_1-f_2}{\hbar},
\end{align}
for various dynamical and background fields. The average ``$r$'' combinations are understood as the ``physical'' macroscopic fields, while the difference ``$a$'' combinations as the stochastic noise associated with them. The SK effective action $S[B_{r},B_{a}]$ for diffusion can be expressed in terms of the global symmetry invariants $B_{r,a\mu}$.

\paragraph*{Diagonal shift symmetry:}

In addition to the global U(1) symmetries above, we impose a local time-independent diagonal shift symmetry between the two phase fields
\begin{equation}
    \varphi_{1,2} \to \varphi_{1,2} 
    + \lambda,
    \qquad \dow_t\lambda = 0,
    \label{eq:shift-SK}
\end{equation}
which acts uniformly on the two legs of the SK contour. This symmetry ensures that the SK-EFT describes the phase where the global U(1) symmetry of the microscopic theory is spontaneously unbroken; see~\cite{Glorioso:2018wxw}.
It is easy to see that the difference combination $B_{a\mu}$ is invariant under the diagonal shift symmetry, while the average combination $B_{r\mu}$ transforms as 
\begin{equation}
    B_{r\mu} \to B_{r\mu} + \dow_\mu\lambda.
\end{equation}
Furthermore, due to the time-independent nature of this symmetry, the time-component $B_{rt}$ of the average combination is also invariant. This is identified as the local chemical potential $\mu\equiv B_{rt}$ in the theory of diffusion.

\paragraph*{Schwinger-Keldysh generating functional:}

The fundamental object of interest in non-equilibrium field theory is the SK generating functional $\cZ[A_r,A_a]$, prescribed as a functional of the two sets of background fields $A_{r,a\mu}$. It can be used to compute the retarded and symmetric correlation functions of $J^\mu$ using the variational formulae 
\begin{align}
    G^\rmR_{J^\mu J^\nu} 
    &= \bfrac{\delta}{\delta A_{r\nu}}
    \bfrac{-i\delta}{\delta A_{a\mu}}
    \ln{\cal Z}[A_r,A_a]
    \bigg|_{A_r,A_a=0}, \nn\\
    G^\rmS_{J^\mu J^\nu} 
    &= \bfrac{-i\delta}{\delta A_{a\mu}}
    \bfrac{-i\delta}{\delta A_{a\nu}}
    \ln{\cal Z}[A_r,A_a]
    \bigg|_{A_r,A_a=0},
    \label{eq:corr-from-gf}
\end{align}
and similarly for higher-point functions.
The generating functional is obtained by performing a path integral of the effective action $S[B_{r},B_{a}]$ over the dynamical fields $\varphi_{r,a}$, as in\footnote{Note that we have absorbed the conventional factor of $1/\hbar$ in the exponent of the path integral into the definition of $S$. This enables the path integral to have a smooth $\hbar\to 0$ limit.} 
\begin{equation}
    {\cal Z}[A_r,A_a]
    = \int\cD\varphi_r\,\cD\varphi_a
    \exp\Big( iS[B_{r},B_{a}]\Big).
\end{equation}

\paragraph*{Unitarity constraints:}

The SK generating functional is required to satisfy the following three unitarity constraints
\begin{equation}
    \cZ[A_r,A_a = 0] = 1, \qquad 
    \cZ[A_r,-A_a] = \cZ^*[A_r,A_a], \qquad 
    \Re\cZ[A_r,A_a] \leq 0,
    \label{eq:SK-conditions-pf}
\end{equation}
arising from generic properties of quantum field theories defined on a closed-time contour.
More details regarding the underlying physics can be found in the review of~\cite{Glorioso:2018wxw}. These constraints can naturally be extended to the effective action as
\begin{equation}
    S[B_r,B_a = 0] = 0, \qquad 
    S[B_r,-B_a] 
    = -S^*[B_r,B_a], \qquad 
    \Im S[B_r,B_a] \geq 0.
    \label{eq:SK-conditions}
\end{equation}
We can arrange $S$ as a series in powers of $B_{a}$, in which case the three conditions mean that: $S$ must at least be linear in $B_{a}$, the terms with even-powers of $B_{a}$ must be imaginary, and the total imaginary part of the effective action must be non-negative.

\paragraph*{Classical equation of motion:} The classical equation of motion for diffusion can be obtained by extremising the effective action with respect to the dynamical fields
\begin{align}
    \frac{\delta S}{\delta\varphi_a}\bigg|_{A_a=0,A_r = A} = 0, \qquad 
    \frac{\delta S}{\delta\varphi_r}\bigg|_{A_a=0,A_r = A} = 0.
\end{align}
Here we have turned off the difference/noise combination of the background gauge fields $A_{a\mu}$, while identified the average combination $A_{r\mu}$ with the classical background gauge field $A_{\mu}$.
Since the effective action is at least linear in noise fields, it follows that the second equation above identically sets the dynamical noise field $\varphi_a$ to zero. On the other hand, the first equation leads to the U(1) conservation equation
\begin{align}\label{eq:classical-EOM}
    \dow_\mu J^\mu = 0, \qquad 
    J^\mu = \frac{\delta S}{\delta A_{a\mu}}\bigg|_{A_a=0,A_r = A} .
\end{align}

\subsection{KMS symmetry}
\label{sec:KMS}

\begin{table}[t]
  \centering
  \begin{tabular}{c|ccc|c|c|c}
    \toprule
    & C & P & T & PT & CT & CPT \\
    \midrule
    $t$, $\dow_t$  & $+$ & $+$ & $-$ & $-$ & $-$ & $-$ \\
    $x^i$, $\partial_i$ & $+$ & $-$ & $+$ & $-$ 
    & $+$ & $-$ \\
    \midrule

    $\varphi$ & $-$ & $+$ & $-$ & $-$ & $+$ & $+$ \\
    $\mu$ & $-$ & $+$ & $+$ & $+$ & $-$ & $-$ \\

    \midrule
    
    $J^t$, $A_t$, $B_t$ & $-$ & $+$ & $+$ & $+$ & $-$ & $-$ \\
    $J^i$, $A_i$, $B_i$ & $-$ & $-$ & $-$ & $+$ & $+$ & $-$ \\
    \bottomrule
  \end{tabular}
  \caption{Eigenvalues under parity (P), time-reversal (T), and charge conjugation (C)
    of various quantities. The Schwinger-Keldysh double copies, in ``$1/2$'' or ``$r/a$'' basis, behave the same as their unlabelled
    versions.\label{tab:CPT}}
\end{table}

The SK conditions in \cref{eq:SK-conditions-pf} need to be satisfied for arbitrary non-equilibrium field theories defined on a closed-time contour. However, to describe correlation functions in a thermal state, the SK generating functional $\cZ[A_r,A_a]$ is further required to satisfy a discrete dynamical KMS (Kubo-Martin-Schwinger) symmetry. The KMS symmetry is responsible for implementing the fluctuation-dissipation relations between symmetric and retarded correlation functions
\begin{equation}
    G^\rmS_{J^\mu J^\nu}(\omega,\bk)
    = \hbar \coth\bfrac{\hbar\beta\omega}{2}
    \Im G^\rmR_{J^\mu J^\nu}(\omega,\bk),
    \label{eq:fdt}
\end{equation}
as well as their higher-point generalisations~\cite{Wang:1998wg}. The KMS symmetry is defined through a transformation of the background fields
\begin{align}
    \rmK A_{1\mu}(t,\bx)
    &= \Theta A_{1\mu}(t - i\theta,\bx), \nn\\
    \rmK A_{2\mu}(t,\bx)
    &= \Theta A_{2\mu}(t + i\hbar\beta  - i\theta,\bx),
    \label{eq:KMS-full}
\end{align}
where $\theta$ is an arbitrary constant parametrising a translations in imaginary time. Whereas, $\Theta$ represents some discrete anti-unitary symmetry of the microscopic theory involving time-reversal transformation (such as time-reversal T, spacetime reversal PT, or variants involving charge conjugation, CT or CPT). See \cref{tab:CPT}. We can check that KMS is a $\bbZ_2$ transformation, $\rmK^2=1$. More details can be found in textbook treatments of thermal field theory, e.g. in~\cite{Bellac:2011kqa}, or in the more recent review of~\cite{Glorioso:2018wxw}. 

In the SK-EFT for diffusion, the action of the KMS symmetry is naturally extended to the dynamical fields as
\begin{align}
    \rmK \varphi_{1}(t,\bx)
    &= \Theta\varphi_{1}(t - i\theta,\bx), \nn\\
    \rmK \varphi_{2}(t,\bx)
    &= \Theta \varphi_{2}(t + i\hbar\beta  - i\theta,\bx),
    \label{eq:KMS-diff}
\end{align}
This is often called the dynamical KMS symmetry. The action of the KMS symmetry on the background fields in \cref{eq:KMS-full} is universal and can be derived from general principles of thermal field theory. On the other hand, the dynamical KMS symmetry in \cref{eq:KMS-diff}, i.e. how KMS transformations act on the dynamical fields, is part of the specifications of the model. In particular, there may be other consistent SK-EFT models where the KMS symmetry acts differently on the dynamical fields; see e.g.~\cite{Jain:2023obu}.

Since the KMS symmetry acts differently on the ``1'' and ``2'' copies of the fields, its action on the SK-EFT---in particular, on the average-difference combinations of fields---is highly non-local in time. However, working perturbatively in derivatives or $\hbar$, the action of the KMS symmetry can be made local. To this end, let us employ a Taylor expansion in time to represent the KMS transformation in \cref{eq:KMS-full,eq:KMS-diff} as
\begin{align}
    \rmK f_{1}
    &= \tilde\Theta \Big( \E{i\hbar\beta/2\dow_t}  f_{1} \Big),\nn\\
    \rmK f_{2}
    &=\tilde\Theta \Big( \E{-i\hbar\beta/2\dow_t}  f_{2}\Big),
\end{align}
where $f$ is a placeholder for $A_\mu$, $\varphi$, or $B_{\mu}$, while $\tilde\Theta \equiv \E{i(\hbar\beta/2-\theta)}\Theta = \Theta\E{-i(\hbar\beta/2-\theta)}$. Passing onto the average-difference basis, this leads to
\begin{align}
    \rmK f_{r} &= \tilde\Theta \lB
    \cos\bfrac{\hbar\beta\dow_t}{2} f_{r} 
    + \frac{i\hbar}{2} \sin\bfrac{\hbar\beta\dow_t}{2}
    f_{a} \rB \nn\\
    &= \tilde\Theta f_r 
    + \frac{\hbar^2}{4}\tilde\Theta 
    \lb 
    i\beta\dow_tf_{a}
    - \frac{\beta^2}{2}\dow_t^2 f_r \rb
    + \cO(\hbar^4), \nn\\
    \rmK f_{a} 
    &= \tilde\Theta \lB 
    \cos\bfrac{\hbar\beta\dow_t}{2} f_{a} 
    + \frac{2i}{\hbar} \sin\bfrac{\hbar\beta\dow_t}{2} f_{r} \rB \nn\\
    &= 
    \tilde\Theta \Big(
    f_{a} 
    + i\beta\dow_tf_{r} \Big)
    - \frac{\hbar^2\beta^2}{8} \tilde\Theta \lb
    \dow_t^2 f_{a} 
    + \frac{i\beta}{3} \dow_t^3f_{r} \rb 
    + \cO(\hbar^4).
    \label{eq:KMS-CS}
\end{align}
This representation will be extremely useful in the next section while constructing the effective action for quantum fluctuations in hydrodynamics.

\subsection{Effective action in $\hbar\to 0$ limit}
\label{sec:classical-action}

The dynamical KMS transformations in \cref{eq:KMS-CS} become quite simple in $\hbar\to 0$ limit. This ``classical'' version of the KMS symmetry is relevant for classical stochastic field theories valid at small frequencies $\omega \ll T_0/\hbar$, where quantum effects are suppressed. This condition precisely yields the classical 2-point fluctuation-dissipation theorem
\begin{equation}
    G^\rmS_{J^\mu J^\nu}(\omega,\bk)
    = \frac{2}{\beta\omega}
    \Im G^\rmR_{J^\mu J^\nu}(\omega,\bk)
    + \cO(\hbar).
    \label{eq:fdt-cl}
\end{equation}
Similar higher-point relations can be derived by using this symmetry at higher non-linear orders in background fields.

As an example, the SK effective action for diffusion in the limit $\hbar\to 0$ is given as~\cite{Crossley:2015evo,Chen-Lin:2018kfl,Jain:2020zhu}
\begin{align}
    S_{\text{cl}}
    &= \int \df t\,\df^d\bx
    \lb
    B_{at} n(B_{rt})
    - \sigma(B_{rt}) B_a^i \dow_t B_{ri}
    + \frac{i}{\beta} \sigma(B_{rt})  B_{a}^iB_{ai} 
    + \cO(\hbar)
    \rb,
    \label{eq:action-SK}   
\end{align}
where we have identified the charge density $n(\mu)$ and the conductivity $\sigma(\mu)$ as functions of $\mu = B_{rt}$. 
The action is manifestly invariant under the global U(1) symmetries and the time-independent diagonal shift symmetry. It trivially obeys the first two SK unitarity constraints in \cref{eq:SK-conditions}. The third condition imposes that the conductivity $\sigma$ is a non-negative function of $\mu$, i.e.
\begin{align}
    \sigma(\mu) \geq 0.
\end{align}

Finally, let us look at the dynamical KMS symmetry. The KMS transformation of the thermodynamic density term in \cref{eq:action-SK}  gives
\begin{align}
    \rmK \Big( B_{at} n(B_{rt})\Big)
    &= \tilde\Theta \Big( 
    \lb B_{at} + i\beta\dow_t B_{rt} \rb 
    n(B_{rt})
    \Big)
    + \cO(\hbar) \nn\\
    &= \tilde\Theta \Big( 
    B_{at}n(B_{rt}) 
    \Big)
    - \dow_t \lb i\beta\, \tilde\Theta p(B_{rt}) \rb
    + \cO(\hbar),
\end{align}
where $p$ is the thermodynamic pressure defined such that $p'(\mu) = n(\mu)$. For $\Theta=\rm{T, PT}$, the $\tilde\Theta$ operator merely implements a change of coordinates $t,\bx$ in the action. We thus infer that the density term in \cref{eq:action-SK} is invariant under KMS transformations up to a total-derivative boundary contribution that can be ignored. The same argument also applies for $\Theta=\rm{CT, CPT}$, except that $n(\mu)$ must be an odd function of $\mu$.

On the other hand, for the conductivity terms in \cref{eq:action-SK},  we note that
\begin{align}
    \rmK\Big( {-}\sigma(B_{rt}) B_a^i \dow_t B_{ri}\Big) 
    &= \tilde\Theta\Big( \sigma(B_{rt}) \lb B_a^i + i\beta\dow_t B_{r}^i\rb \dow_t B_{ri}\Big) 
    + \cO(\hbar) \nn\\
    &= \tilde\Theta\Big( \sigma(B_{rt}) B_a^i \dow_t B_{ri}\Big) 
    + i\beta\,\tilde\Theta\Big( \sigma(B_{rt}) \dow_t B_{r}^i \dow_t B_{ri}\Big)
    + \cO(\hbar), \nn\\
    \rmK\lb \frac{i}{\beta} \sigma(B_{rt})  B_{a}^iB_{ai} \rb
    &= \tilde\Theta\lb \frac{i}{\beta} \sigma(B_{rt})  
    \lb B_a^i + i\beta\dow_t B_{r}^i\rb
    \lb B_{ai} + i\beta\dow_t B_{ri}\rb \rb 
    + \cO(\hbar) \nn\\
    &=  \tilde\Theta\lb \frac{i}{\beta} \sigma(B_{rt})B_a^i B_{ai} \rb
    - 2 \tilde\Theta\Big( \sigma(B_{rt})  B_a^i \dow_t B_{ri}\Big) \nn\\
    &\qquad 
    - i\beta \tilde\Theta\Big( \sigma(B_{rt})  
    \dow_t B_{r}^i \dow_t B_{ri}\Big)
    + \cO(\hbar).
\end{align}
Therefore, these terms also map to themselves up to the $\tilde\Theta$ transformation. As before, the conductivity terms are identically KMS-invariant for $\Theta=\rm{T, PT}$, while for $\Theta=\rm{CT,CPT}$ we further require that $\sigma(\mu)$ is an even function of $\mu$.

It is often useful to cast the SK effective action explicitly in terms of $\mu$. We find
\begin{align}
    S_{\text{cl}}
    &= \int \df t\,\df^d\bx
    \lb
    B_{at} n(\mu)
    - \sigma(\mu) B_a^i 
    \Big( \dow_i\mu - E_i\Big)
    + \frac{i}{\beta} \sigma(\mu)  B_{a}^iB_{ai} 
    + \cO(\hbar)
    \rb,
\end{align}
where $E_i = \dow_i A_{rt}-\dow_t A_{ri} $ is the background electric field.

\section{Quantum-complete Schwinger-Keldysh effective action}
\label{sec:quantum-action}

In \cref{sec:SK-formalism}, we obtained the SK effective action for diffusion in $\hbar\to 0$ limit. Our aim in this section is to complete this effective action with appropriate quantum corrections such that it satisfies the full KMS symmetry in \cref{eq:KMS-CS} up to arbitrary orders in $\hbar$. We will start with the free quadratic version of the theory in \cref{sec:quantum-action-free}, where it is easy to guess the correct form of the effective action based on FDT. We then derive the complete interacting non-linear effective action in \cref{sec:quantum-action-full}.

\subsection{Free effective action}
\label{sec:quantum-action-free}

To obtain the quantum-complete version of SK effective action, it is easiest to start with the free theory, with $n(\mu)=\chi_0\mu$ and $\sigma(\mu)=\sigma_0$, where $\chi_0$ is the constant thermodynamic charge susceptibility and $\sigma_0\geq 0$ is the constant conductivity. In this case, the classical effective action reduces to
\begin{align}
    S_{\text{cl,free}}
    &= \int \df t\,\df^d\bx \lb 
    \chi_0 B_{rt} B_{at}
    - \sigma_0 B^i_a \dow_t B_{ri} 
    + \frac{i}{\beta} \sigma_0 B^i_a B_{ai}
    + \cO(\hbar) \rb.
\end{align}
At the level of two-point functions, obtaining the quantum version of FDT in \cref{eq:fdt} from the classical one in \cref{eq:fdt-cl} merely corresponds to promoting the noise strength from $2/(\beta\omega)$ to $\hbar\coth(\hbar\beta\omega/2)$. Correspondingly, we can guess the quantum-complete version of the free effective action by appropriately promoting the noise term
\begin{align}
    S_{\text{free}}
    &= \int \df t\,\df^d\bx \lb n_0 B_{at} + \chi_0 B_{rt} B_{at}
    - \sigma_0 B^i_a \dow_t B_{ri} 
    + \frac{i}{\beta} \sigma_0 B^i_a \fD B_{ai}
    + \cO(\hbar) \rb,
    \label{eq:quantum-action-free}
\end{align}
where we have defined the differential operator $\fD$ as
\begin{align}
    \fD &= \frac{\hbar\beta\dow_t}{2}
    \cot\bfrac{\hbar\beta\dow_t}{2} \nn\\
    &= 1 - \frac{\hbar^2\beta^2}{12}\dow_t^2
    - \frac{\hbar^4\beta^4}{720}\dow_t^4
    + \ldots.
\end{align}
Note that the operator $\fD$, and thus the effective action, is completely local within the derivative expansion of effective field theories. One may check that \cref{eq:quantum-action-free} continues to obey all the SK requirements: global U(1) symmetries, diagonal shift symmetry, as well as the three SK unitarity constraints.

\paragraph*{Verifying the KMS symmetry:}

Let us  verify that the effective action in \cref{eq:quantum-action-free} is indeed invariant under full KMS transformations in \cref{eq:KMS-CS}. To this end, it is useful to introduce a short-hand notation
\begin{align}
    \sfC = \cos\bfrac{\hbar\beta\dow_t}{2}, \qquad 
    \sfS = \sin\bfrac{\hbar\beta\dow_t}{2}.
\end{align}
In terms of these, $\fD = \hbar\beta/2\,\dow_t \sfC/\sfS$.
Further, note that $\E{\lambda\dow_t} \big( fg\big) = \lb\E{\lambda\dow_t} f\rb \lb\E{\lambda\dow_t} g \rb$ for arbitrary functions $f$ and $g$.
It follows that 
\begin{align}
    \sfC \big( fg\big) &= \sfC f\,\sfC g - \sfS f\,\sfS g, \nn\\
    \sfS \big( fg\big) &= \sfC f\,\sfS g + \sfS f\,\sfC g .
\end{align}

The KMS transformation of the $\chi_0$ term in \cref{eq:quantum-action-free} gives
\begin{align}
    \rmK\Big( B_{rt}B_{at} \Big)
    &= \tilde\Theta\lB 
    \lb \sfC B_{rt} 
    + \frac{i\hbar}{2} \sfS B_{at} \rb
    \lb \sfC B_{at}
    + \frac{2i}{\hbar} \sfS B_{rt}\rb
    \rB \nn\\
    &= \tilde\Theta\lB 
    \sfC B_{rt} \sfC B_{at} 
    + \frac{2i}{\hbar} \sfC B_{rt} \sfS B_{rt}
    + \frac{i\hbar}{2} \sfS B_{at} \sfC B_{at}
    - \sfS B_{at} \sfS B_{rt}
    \rB \nn\\
    &= \tilde\Theta\lB 
    \sfC  \Big(B_{rt}B_{at} \Big)
    + \sfS \lb \frac{i}{\hbar} B_{rt}^2\rb
    + \sfS \lb \frac{i\hbar}{4} B_{at}^2\rb
    \rB.
\end{align}
Since the Taylor expansion of $\sfS$ starts from $\hbar\beta/2\,\dow_t + \ldots$, it follows that $\sfS(f) = \dow_t(\ldots)$. Therefore, the last two terms above drop out as total-derivative boundary contributions in the effective action. On the other hand, the Taylor expansion of $\sfC$ starts with $1 - \hbar^2\beta^2/8\,\dow_t^2 + \ldots$, therefore we have $\sfC(f) = f + \dow_t(\ldots)$. Using this we find immediately see that
\begin{align}
    \rmK\Big( B_{rt}B_{at} \Big)
    &= \tilde\Theta\Big(B_{rt} B_{at} \Big) 
    + \dow_t(\ldots).
\end{align}
Since $B_{rt}B_{at}$ is even under charge conjugation, it follows that the $\chi_0$ term in \cref{eq:quantum-action-free} is invariant under KMS for any choice of $\Theta$.

% \begin{align}
%     \rmK B_{at}
%     &= \tilde\Theta B_{at}
%     + \dow_t(\ldots).
% \end{align}
% Therefore, the $n_0$ term is invariant under the full KMS symmetry for $\Theta=\rm{T,PT}$. However, for $\Theta=\rm{CT,CPT}$, the term changes by an overall sign and we are forced to set $n_0=0$.

% For the $n_0$ term we can show that
% \begin{align}
%     \rmK B_{at}
%     &= \tilde\Theta\bigg[
%     \sfC B_{at}
%     + \sfS \lb\frac{2i}{\hbar} B_{rt} \rb
%     \bigg].
% \end{align}

Let us now look at the two $\sigma_0$ terms. In this case, we have
\begin{align}
    \rmK\Big( {-} B^i_a \dow_t B_{ri} \Big)
    &= \tilde\Theta\lB 
    \lb \sfC B_{a}^i
    + \frac{2i}{\hbar} \sfS B_{r}^i\rb
    \lb \sfC \dow_t B_{ri} 
    + \frac{i\hbar}{2} \sfS\dow_t B_{ai}\rb
    \rB \nn\\
    &= \tilde\Theta\lB 
    \sfC B_{a}^i\sfC \dow_t B_{ri}
    - \sfS B_{r}^i\sfS\dow_t B_{ai}
    + \frac{i\hbar}{2}\sfC B_{a}^i \sfS\dow_t B_{ai}
    + \frac{2i}{\hbar} \sfS B_{r}^i \sfC \dow_t B_{ri} 
    \rB \nn\\
    &= \tilde\Theta\bigg[
    \sfC B_{a}^i\sfC \dow_t B_{ri}
    - \sfS B_{r}^i\sfS\dow_t B_{ai}
    - \frac{i\hbar}{2} \sfS B_{a}^i \sfC\dow_t B_{ai}
    + \frac{2i}{\hbar} \sfS B_{r}^i \sfC \dow_t B_{ri} 
     \nn\\
    &\qquad\qquad 
    + \sfS 
    \lb \frac{i\hbar}{2} B_{a}^i\dow_t B_{ai} \rb \bigg], \nn\\
    \rmK \lb \frac{i}{\beta} B^i_a \fD B_{ai} \rb
    &= \tilde\Theta \lB \frac{i\hbar}{2} 
    \lb \sfC B_{a}^i + \frac{2i}{\hbar} \sfS B^i_{r}\rb
    \frac{\sfC}{\sfS} 
    \lb \sfC \dow_t B_{ai} + \frac{2i}{\hbar} \sfS \dow_t B_{ri}\rb \rB \nn\\
    &= \tilde\Theta \lB 
    \frac{i\hbar}{2}  
    \sfC B_{a}^i\frac{\sfC^2}{\sfS} \dow_t B_{ai} 
    -  \sfC B_{a}^i \sfC \dow_t B_{ri}
    -  \sfS B^i_{r} \frac{\sfC^2}{\sfS}  \dow_t B_{ai} 
    - \frac{2i}{\hbar}  \sfS B^i_{r} \sfC \dow_t B_{ri}
    \rB \nn\\
    &= \tilde\Theta \bigg[
    \frac{i\hbar}{2}  
    \sfC B_{a}^i\frac{\sfC^2}{\sfS} \dow_t B_{ai} 
    -  \sfC B_{a}^i \sfC \dow_t B_{ri}
    +  \sfC B^i_{r} \sfC  \dow_t B_{ai} 
    - \frac{2i}{\hbar}  \sfS B^i_{r} \sfC \dow_t B_{ri} \nn\\
    &\qquad\qquad 
    - \sfS \lb  B^i_{r} \frac{\sfC}{\sfS}  \dow_t B_{ai} \rb
    \bigg].
\end{align}
Adding the two results, we find
\begin{align}
    \rmK\lb {-}  B^i_a \dow_t B_{ri}
    + \frac{i}{\beta}  B^i_a \fD B_{ai}\rb
    &= \tilde\Theta \bigg[
    \sfC \Big( B^i_{r} \dow_t B_{ai} \Big)
    + \sfC \lb \frac{i\hbar}{2}   B_{a}^i\frac{\sfC}{\sfS} \dow_t B_{ai} \rb
    \bigg]
    + \dow_t(\ldots) \nn\\
    &= \tilde\Theta \bigg[
     B^i_{r} \dow_t B_{ai}
    + \frac{i}{\beta}   B_{a}^i\fD B_{ai}
    \bigg]
    + \dow_t(\ldots) \nn\\
    &= \tilde\Theta \bigg[
    {-} B_{ai} \dow_t B^i_{r}
    + \frac{i}{\beta}   B_{a}^i\fD B_{ai}
    \bigg]
    + \dow_t(\ldots).
\end{align}
Therefore, the $\sigma_0$ terms are also invariant under KMS for any choice of $\Theta$.

\subsection{Non-linear effective action}
\label{sec:quantum-action-full}

The full interacting version of the action is slightly trickier to guess. One can convince themselves that simply promoting the coefficients in \cref{eq:quantum-action-free} to be generic functions of $B_{rt}$ does not work. 
% For example, if we were to take the next correction in charge density, $n\sim \half\chi_1\delta\mu^2$, the respective KMS transformation yields
% \begin{align}
%     \rmK\Big( B_{rt}^2B_{at} \Big)
%     &= \tilde\Theta\lB 
%     \lb \sfC B_{rt} 
%     + \frac{i\hbar}{2} \sfS B_{at} \rb^2
%     \lb \sfC B_{at}
%     + \frac{2i}{\hbar} \sfS B_{rt}\rb
%     \rB \nn\\
%     % 
%     &= \tilde\Theta\lB 
%     \chi_0 \sfC B_{rt} \sfC B_{at} 
%     + \frac{2i}{\hbar} \chi_0 \sfC B_{rt} \sfS B_{rt}
%     + \frac{i\hbar}{2}  \chi_0 \sfS B_{at} \sfC B_{at}
%     - \chi_0 \sfS B_{at} \sfS B_{rt}
%     \rB \nn\\
%     % 
%     &= \tilde\Theta\lB 
%     \sfC  \Big(\chi_0B_{rt}B_{at} \Big)
%     + \sfS \lb \frac{i}{\hbar} \chi_0 B_{rt}^2\rb
%     + \sfS \lb \frac{i\hbar}{4}  \chi_0 B_{at}^2\rb
%     \rB \nn\\
%     % 
%     &= \tilde\Theta\Big(\chi_0B_{rt} B_{at} \Big) 
%     + \dow_t(\ldots).
% \end{align}
However, we can use the following trick. For some function $F(f)$ defined over a single copy of fields, let us define the average-difference combinations
\begin{align}
    F(f)_r = \frac{F(f_1) + F(f_2)}{2} 
    &= \cosh\lb\frac{\hbar}{2} f_a \frac{\dow}{\dow f_r}\rb F(f_r) \nn\\
    &= F(f_r)
    + \frac{\hbar^2}{8} f_a^2 F''(f_r)
    + \ldots, \nn\\
    F(f)_a = \frac{F(f_1) - F(f_2)}{\hbar}
    &= \frac{2}{\hbar}\sinh\lb\frac{\hbar}{2} f_a \frac{\dow}{\dow f_r}\rb F(f_r) \nn\\
    &= f_a F'(f_r)
    + \frac{\hbar^2}{24} f_{a}^3 F'''(f_r)
    + \ldots.
\end{align}
Note also that
\begin{align}
    \Big( F(f)G(f) \Big)_r 
    &= F(f)_r G(f)_r + \frac{\hbar^2}{4}F(f)_aG(f)_a, \nn\\
    \Big( F(f)G(f) \Big)_a
    &= F(f)_r G(f)_a + F(f)_aG(f)_r.
\end{align}
It is easy to check that the KMS transformations of $F(f)_{r,a}$ take the same form in \cref{eq:KMS-CS}, i.e.
\begin{align}
    \rmK F(f)_{r} &= \tilde\Theta \lB
    \cos\bfrac{\hbar\beta\dow_t}{2} F(f)_{r} 
    + \frac{i\hbar}{2} \sin\bfrac{\hbar\beta\dow_t}{2}
    F(f)_{a} \rB, \nn\\
    \rmK F(f)_{a} 
    &= \tilde\Theta \lB 
    \cos\bfrac{\hbar\beta\dow_t}{2} F(f)_{a} 
    + \frac{2i}{\hbar} \sin\bfrac{\hbar\beta\dow_t}{2} F(f)_{r} \rB.
    \label{eq:KMS-general}
\end{align}
Therefore, we can recycle the derivations in \cref{sec:quantum-action-free} to show that the combinations 
\begin{align}
    \fL_1[F(B)]
    &= F(B)_a,  \nn\\
    \fL_2[F(B),G(B)] 
    &= i F(B)_a G(B)_a 
    - \frac{1}{\hbar} 
    \lb F(B)_a \frac{\sfS}{\sfC} G(B)_r 
    + G(B)_a \frac{\sfS}{\sfC} F(B)_r \rb,
    \label{eq:KMS-combos}
\end{align}
is invariant under KMS for any functions $F$ and $G$. 
In fact, the free terms in the previous subsection can be viewed as special cases of these combinations
\begin{align}
    \fL_1\lB \half \chi_0 B_t^2\rB
    &= \chi_0  B_{at} B_{rt}, \nn\\
    \fL_2\lB \frac{1}{\beta}\sigma_0 B^i, \fD B_{i}\rB
    &= \frac{i}{\beta}\sigma_0 B_{a}^i\fD B_{ai} 
    - \frac{1}{\hbar\beta}\sigma_0 \lb 
    B_{a}^i \frac{\sfS}{\sfC} \fD B_{ri} 
    + \fD B^i_a \frac{\sfS}{\sfC} B_{ai} \rb \nn\\
    &=
    \sigma_0 B_{a}^i \lb \frac{i}{\beta} \fD B_{ai} 
    - \dow_t B_{ri} \rb
    + \dow_t(\ldots).
    \label{eq:KMS-combo-free}
\end{align}
Note that $g(\fD f) = (\fD g) f + \dow_t(\ldots)$. Having been constructed in terms of the invariants $B_{r,a\mu}$, the combinations in \cref{eq:KMS-combos} are manifestly invariant the global U(1) symmetries. They also trivially satisfy the first two unitarity constraints in \cref{eq:SK-conditions}. 
However, we will need to verify the third unitarity constraint and the diagonal shift symmetry on case-by-case basis.

\paragraph*{Thermodynamic density term:}

Having identified the invariants in \cref{eq:KMS-combos}, lifting the thermodynamic density term to the quantum level is quite straightforward. We can simply take
\begin{align}
    \fL_1[p(B_{t})] &= p(B_t)_a \nn\\
    &= B_{at} n(B_{rt})
    + \frac{\hbar^2}{24} B_{at}^3 n''(B_{rt})
    + \ldots.
    \label{eq:thermo-Lagrangian}
\end{align}
Since this expression is exclusively constructed out of $B_{r,at}$ and has no imaginary part, it trivially satisfies the third unitarity constraint and the diagonal shift symmetry.
We see that quantum corrections generate terms at higher orders in noise fields.

\paragraph*{Dissipative conductivity terms:} 

Generalising the conductivity term to nonzero $\hbar$ is a bit more subtle. Let us try the straightforward generalisation of \cref{eq:KMS-combo-free}, leading to
\begin{align}
    \fL_2\lB \frac{1}{\beta}\sigma(B_t) B^i, \fD B_{i}\rB
    &= \frac{i}{\beta} \lb \sigma(B_t) B^i\rb_a \fD B_{ai} 
    - \frac{1}{\hbar\beta} 
    \lb \lb \sigma(B_t) B^i\rb_a  \frac{\sfS}{\sfC} \fD B_{ri} 
    + \fD B^i_a \frac{\sfS}{\sfC} \lb\sigma(B_t) B^i\rb_r \rb \nn\\
    &= \frac{i}{\beta} \lb \sigma(B_t) B^i\rb_a \fD B_{ai} 
    - \frac{1}{2} \lb \sigma(B_t) B^i\rb_a  \dow_t B_{ri} 
    - \half B^i_a \dow_t \lb\sigma(B_t) B^i\rb_r 
    + \dow_t(\ldots) \nn\\
    &= \sigma(B_t)_r B^i_a 
    \lb \frac{i}{\beta} \fD B_{ai} 
    - \dow_t B_{ri}
    \rb 
    + \frac{i}{\beta} \sigma(B_t)_a B^i_r \fD B_{ai} 
    - \frac{\hbar^2}{16} \dow_t\sigma(B_t)_a\, B^i_a B_{ai} \nn\\
    &\qquad 
    - \frac{1}{2} \dow_t \sigma(B_t)_r\,B^i_a B^i_r
    + \frac{1}{4} \dow_t \sigma(B_t)_a\,B^i_r B_{ri}
    + \dow_t(\ldots).
    \label{eq:incomplete-dissipative}
\end{align}
Here the terms in the first line are fine: they generate quantum corrections to the SK effective action starting at quadratic and cubic orders in noise fields. However, the remaining terms in the last line generate corrections at the linear order in noise fields and thus modify the classical conserved current of the diffusion model in \cref{eq:classical-EOM}. Therefore, while fully consistent with the KMS symmetry at finite $\hbar$, the dissipative Lagrangian in \cref{eq:incomplete-dissipative} does not describe quantum corrections to the original diffusion model we wanted to study.\footnote{I would like to thank Blaise Gout\'eraux for discussion on this point.}
There is also a second problem with \cref{eq:incomplete-dissipative}: it is not invariant under the diagonal shift symmetry in \cref{eq:shift-SK}.\footnote{The two issues are not connected. It is possible to generate variants of dissipative Lagrangian that respect the diagonal shift symmetry, but still modify the classical conserved current. Firstly, one may verify that the following combination is KMS-invariant for any functions $F$, $G$, and $H$, i.e.
\begin{align}
    \fL_3[F(B),G(B),H(B)] = \lb H(B)_r F(B)_a 
    - \frac{i\hbar}{2} H(B)_a \frac{\sfS}{\sfC} F(B)_a
    \rb \lb iG(B)_{a} - \frac{2}{\hbar} \frac{\sfS}{\sfC} G(B)_r \rb
    + (F\leftrightarrow G).
\end{align}
In fact, $\fL_2$ is a special case of this invariant for $H(B) = 1/2$. Using this, one can construct
\begin{align}
    \fL_3\lB B^i,\fD B_{i},\frac{1}{2\beta}\sigma(B_{t})\rB
    &= \frac{i}{\beta}\sigma(B_t)_r B^i_a \fD B_{ai}
    - \sigma(B_t)_r B^i_a \dow_t B_{ri} 
    + \frac{\hbar}{4\beta} \sigma(B_t)_a \frac{\sfS}{\sfC} B^i_a\fD B_{ai} 
    - \frac{\hbar^2}{16} \dow_t\sigma(B_t)_a B_{ai}B^i_{a}  
    \nn\\
    &
    + \frac{i\hbar}{4} \sigma(B_t)_a \lb 
    \dow_t B_{ai}\frac{\sfS}{\sfC} B^i_r 
    + \frac{\sfS}{\sfC} B^i_a \dow_t B_{ri}  \rb
    + \half \sigma(B_t)_r \lb B_{ai} \dow_t B^i_r 
    - \frac{\sfC}{\sfS} \dow_t B_{ai} \frac{\sfS}{\sfC} B^i_r \rb.
\end{align}
Since the Taylor expansion of $\sfS/\sfT$ starts with $\hbar\beta/2\,\dow_t$, this expression is manifestly invariant under the time-independent diagonal shift symmetry. This Lagrangian is also local perturbatively in derivatives. However, it does contain improvements that are linear in noise fields and thus modify the classical conserved current.
}

Fortunately, both of these problems have an easy fix. We include another combination in the dissipative Lagrangian
\begin{align}
    \fL_2\lB \frac{-1}{2\beta} \fD\sigma(B_t), B^i B_i\rB
    &= \frac{-i}{2\beta}  \fD\sigma(B_t)_a \lb B^i B_i\rb_a 
    + \frac{1}{2\hbar\beta} 
    \lb  \fD\sigma(B_t)_a \frac{\sfS}{\sfC} \lb B^i B_i\rb_r 
    + \lb B^i B_i\rb_a \frac{\sfS}{\sfC}  \fD\sigma(B_t)_r \rb \nn\\
    &= -\frac{i}{\beta} \fD \sigma(B_t)_a \lb B^i_a B_{ri}\rb
    - \frac{\hbar^2}{16} \dow_t\sigma(B_t)_a B^i_a B_{ai}
    \nn\\
    &\qquad 
    + \half \dow_t \sigma(B_t)_r\,B^i_a B_{ri}
    - \frac{1}{4} \dow_t\sigma(B_t)_a B^i_r B_{ri}
    + \dow_t(\ldots).
\end{align}
The terms in the second line here cancel with those in \cref{eq:incomplete-dissipative}. Therefore, the complete dissipative Lagrangian is given as
\begin{align}
    &\fL_2\lB \frac{1}{\beta}\sigma(B_t) B^i, \fD B_{i}\rB
    + \fL_2\lB \frac{-1}{2\beta}\sigma(B_t), \fD \lb B^i B_i\rb\rB \nn\\ 
    &= \sigma(B_t)_r B^i_a 
    \lb \frac{i}{\beta} \fD B_{ai} 
    - \dow_t B_{ri}
    \rb 
    + \frac{i}{\beta} B^i_r  \Big( \sigma(B_t)_a \fD B_{ai} 
    - \fD \sigma(B_t)_a B_{ai} \Big)
    - \frac{\hbar^2}{8} \dow_t\sigma(B_t)_a\, B^i_a B_{ai} .
    \label{eq:dissipative-Lagrangian}
\end{align}

The third unitarity constraint in \cref{eq:SK-conditions} requires that the imaginary part of the effective action is non-negative. Truncating at leading order in derivatives, the imaginary part of \cref{eq:dissipative-Lagrangian} is simply
\begin{align}
    \frac{1}{\beta}\sigma(B_t)_r B^i_aB_{ai}
    + \cO(\dow_t^2).
\end{align}
Since $\sigma(\mu)$ is a non-negative function, it follows that $\sigma(B_t)_r = (\sigma(B_{1t})+\sigma(B_{2t}))/2$ is also non-negative. Therefore, the imaginary part of \cref{eq:dissipative-Lagrangian} is non-negative at leading order in derivatives and hence is consistent with the unitarity constraints. Note that effective field theories are only defined within a derivative expansion, so it is sufficient to impose the non-negativity condition at leading order in derivatives.

The dissipative Lagrangian in \cref{eq:dissipative-Lagrangian} is also consistent with the diagonal shift symmetry. Its variation under \cref{eq:shift-SK} is given as
\begin{align}
    \frac{i}{\beta} \dow_i\lambda  \Big( \sigma(B_t)_a \fD B_{ai} 
    - \fD \sigma(B_t)_a B^i_a \Big).
\end{align}
Since $g(\fD f) - (\fD g) f = \dow_t(\ldots)$ and $\dow_t\lambda=0$, this expression evaluates to a total-derivative boundary contribution in the effective action.

\paragraph*{Complete effective action:} Combining \cref{eq:thermo-Lagrangian,eq:dissipative-Lagrangian}, the quantum-complete SK effective action for diffusion can be written as
\begin{align}
    S
    &= \int \df t\,\df^d\bx \bigg[
    p(B_t)_a
    + \sigma(B_t)_r B^i_a 
    \lb \frac{i}{\beta} \fD B_{ai} 
    - \dow_t B_{ri}
    \rb   \nn\\
    &\hspace{8em} 
    + \frac{i}{\beta} B^i_r  \Big( \sigma(B_t)_a \fD B_{ai} 
    - \fD \sigma(B_t)_a B^i_a \Big)
    - \frac{\hbar^2}{8} \dow_t\sigma(B_t)_a\, B^i_a B_{ai}
    \bigg].
    \label{eq:quantum-action}
\end{align}
This compact form of the effective action is quite cryptic. To get some physical intuition, let us expand it to quartic order in noise fields, leading to
\begin{align}
    S
    &= \int \df t\,\df^d\bx \bigg[
    B_{at} n(B_{rt})
    - \sigma(B_{rt}) B^i_a \dow_t B_{ri} \nn\\
    &\qquad \qquad
    + \frac{i}{\beta} \sigma(B_{rt}) B^i_a \fD B_{ai} 
    + \frac{i}{\beta} B_{ri} \bigg( 
    B_{at} \sigma'(B_{rt}) \fD B_{a}^i 
    - B^i_a\, \fD\big(B_{at}\sigma'(B_{rt})\big) \bigg)
    \nn\\
    &\qquad\qquad 
    + \frac{\hbar^2}{24} B_{at}^3 n''(B_{rt}) 
    - \frac{\hbar^2}{8} \sigma''(B_{rt}) B_{at}^2 B^i_a \dow_t B_{ri}
    - \frac{\hbar^2}{8} 
    \dow_t\Big(\sigma'(B_{rt}) B_{at} \Big) B^i_a B_{ai}
    \nn\\
    &\qquad \qquad
    + \frac{i\hbar^2}{8\beta} \sigma''(B_{rt})
    B_{at}^2 B^i_a \fD B_{ai} 
    + \frac{i\hbar^2}{24\beta} B_{ri}  
    \bigg( B_{at}^3 \sigma'''(B_{rt}) \fD B_{a}^i 
    - B^i_a\fD\big( B_{at}^3 \sigma'''(B_{rt})\big) \bigg) \nn\\
    &\qquad \qquad
    + \cO(B_a^5) \bigg].
\end{align}
It can also be expressed in terms of the chemical potential $\mu$ as
\begin{align}
    S
    &= \int \df t\,\df^d\bx \bigg[
    B_{at} n(\mu)
    - \sigma(\mu) B^i_a \lb \dow_i\mu - E_i \rb \nn\\
    &\qquad \qquad
    + \frac{i}{\beta} \sigma(\mu) B^i_a \fD B_{ai} 
    - \frac{i}{\beta} \fD_2\Big( B^i_{a}, B_{at} \sigma'(\mu) \Big)
    \lb \dow_i\mu - E_i \rb
    \nn\\
    &\qquad\qquad 
    + \frac{\hbar^2}{24} B_{at}^3 n''(\mu) 
    - \frac{\hbar^2}{8} \sigma''(\mu) B_{at}^2 B^i_a 
    \lb \dow_i\mu - E_i \rb
    - \frac{\hbar^2}{8} 
    \dow_t\Big(\sigma'(\mu) B_{at} \Big) B^i_a B_{ai}
    \nn\\
    &\qquad \qquad
    + \frac{i\hbar^2}{8\beta} \sigma''(\mu)
    B_{at}^2 B^i_a \fD B_{ai} 
    - \frac{i\hbar^2}{24\beta} 
    \fD_2\Big(B^{i}_a, B_{at}^3 \sigma'''(\mu) \Big)
    \lb \dow_i\mu - E_i \rb
    + \cO(B_a^5) \bigg].
\end{align}
where we have defined the bilinear operator $\fD_2$ via $g\fD f - f\fD g = \dow_t\fD_2(f,g)$.\footnote{We can express $\fD_2$ as a Taylor expansion in derivatives
\begin{align}
    \fD_2(f,g)
    = \sum_{m=1}^{\infty}
    \sum_{n=1}^{2m} 
    (-1)^{m+n}\,\frac{B_{2m}(\hbar\beta)^{2m}}{(2m)!}
    \dow_t^{n-1} f\,\partial_t^{2m-n}g,
\end{align}
where $B_{2m}$ are the Bernoulli's numbers arising from the Taylor expansion of $\cot(\hbar\beta/2\,\dow_t)$.} We will use this effective action in \cref{sec:retarded-correlators} to compute one-loop corrections to the 2-point retarded correlation function of density.

We should emphasise that this construction is not exhaustive: \cref{eq:quantum-action} is not the unique quantum extension of the classical SK effective action for diffusion in \cref{eq:action-SK}. Even purely at the classical level, it is known that the SK effective action admits non-universal ambiguities that do not affect the classical conserved currents; see~\cite{Jain:2020zhu}. Such ambiguities likely also affect the quantum corrections. Generally, these ambiguities only show up at quartic and higher-orders in fields (including both noise and physical fields). The KMS symmetry is strong enough to completely fix the SK effective action up to cubic order in fields. Correspondingly, we believe that the effective action \eqref{eq:quantum-action} is also unambiguous until cubic order in fields. This will be sufficient for our one-loop calculation in \cref{sec:retarded-correlators}.

\section{One-loop retarded correlation functions}
\label{sec:retarded-correlators}

In this section we use the quantum-complete SK effective action for diffusion in \cref{eq:quantum-action} to derive the one-loop correction to the two-point retarded correlation function of density. We begin by setting up a linearised perturbative expansion in fields in \cref{sec:Feynman-rules} and identify the Feynman rules for free propagators and interaction vertices. Some details of the Feynman rules concerning background field insertions have been relegated to \cref{app:feynman-rules-source}. We use these rules to calculate one-loop diagrams contributing to the two-point retarded correlation function in \cref{sec:one-loop}.

\subsection{Feynman rules}
\label{sec:Feynman-rules}

\paragraph*{Free propagators:}

We work around an equilibrium state given by constant chemical potential $\mu = \mu_0$, or equivalently $\varphi_r=\mu_0t$ and $\varphi_a = 0$.
The effective action \eqref{eq:quantum-action} is parametrised by two functions $n(\mu)$ and $\sigma(\mu)$. We can expand these around the equilibrium state as
\begin{align}
    n(\mu) 
    &= n_0 + \chi_0 \delta\mu 
    + \frac12 \chi_1 \delta\mu^2
    + \frac{1}{6} \chi_2 \delta\mu^3
    + \ldots, \nn\\
    \sigma(\mu) 
    &= \sigma_0 
    + \sigma_1 \delta\mu 
    + \half \sigma_2 \delta\mu^2 + \ldots.
\end{align}
Plugging these into \cref{eq:quantum-action}, the free action is given as
\begin{align}
    S_2
    &= \int \df^d t\,\df^d\bx 
    \bigg[ n_0 B_{at} 
    + \chi_0 B_{at} \delta B_{rt}
    - \sigma_0 B^i_a \dow_t B_{ri} 
    + \frac{i}{\beta} \sigma_0 B^i_a \fD B_{ai}
    \bigg],
\end{align}
where $\delta B_{rt} = B_{rt} -\mu_0 = \dow_t\delta\varphi_r + A_{rt}$ and $B_{ri} = \dow_i\delta\varphi_r + A_{ri}$. 

To set up the Feynman diagrammatic expansion, we use solid lines
\tikzsetnextfilename{line-r}%
(\raisebox{2pt}{\tikz[thick]{\draw[] (0,0)--(5mm,0);}}) 
to represent the physical field $\delta\varphi_r$, wavy lines 
\tikzsetnextfilename{can_line-a}%
(\raisebox{1pt}{\tikz[thick]{\draw[aux] (0,0)--(5mm,0);}})
to represent teh noise field $\varphi_a$, single arrows 
\tikzsetnextfilename{can_back-a}%
(\raisebox{-1pt}{\tikz[thick]{\fextin{}{0mm}}}) 
for $iA_{at}$ variations, and double arrows 
\tikzsetnextfilename{can_back-r}%
(\raisebox{-1pt}{\tikz[thick]{\fextin{2}{0mm}}}) 
for $A_{rt}$ variations.
As we are only interested in the correlation functions of density, we will set $A_{r,ai}=0$.

The free part of the Lagrangian gives rise to the following nonzero tree-level propagators
\begin{alignat}{2}
    \tikzsetnextfilename{prop-ra}
    \tikz[thick]{\fline[10mm]{}{aux}{0mm}}
    ~~
    &= \Big\langle\,\delta\varphi_r(p)\,\varphi_a(-p)\Big\rangle_\text{tree}
    &&= \frac{1}{\chi_0\omega}
    \frac{1}{\Delta(p)}
    \equiv \frac{1}{\chi_0\omega}
    \frac{1}{D\bk^2 - i\omega}, \nn\\
    \tikzsetnextfilename{prop-rr}
    \tikz[thick]{\fline[10mm]{}{}{0mm}}
    ~~
    &= \Big\langle\,\delta\varphi_r(p)\delta\varphi_r(-p)\Big\rangle_\text{tree}
    &&= 
    \frac{2}{\chi^2_0\omega^2\beta}
    \frac{\bk^2\sigma_0\fD}{\Delta_\rmS(p)}
    \equiv \frac{2}{\chi^2_0\omega^2\beta}
    \frac{\bk^2\sigma_0\fD}{|D\bk^2-i\omega|^2}  \nn\\
    &&&=
    \frac{\fD}{\chi_0\omega^2\beta} \lb
    \frac{1}{\Delta(p)}
    + \frac{1}{\Delta(-p)} \rb,
    \label{eq:free-props}
\end{alignat}
where $(p)=(\omega,\bk)$ with frequency $\omega$ and wavevector $k^i$. Further $D = \sigma_0/\chi_0$ is diffusivity. Note that $\fD= \hbar\beta\omega/2\,\coth(\hbar\beta\omega/2)$ in Fourier-space.
The symmetric ``$rr$'' propagator has poles in both the upper-half and lower-half complex-$\omega$ planes, while the retarded ``$ra$'' propagator only has poles in the lower-half complex-$\omega$ plane.
Importantly, since the Lagrangian has no terms with only $r$-type fields, the ``$aa$'' noise propagator $\langle\phi_a(p)\phi_a(-p)\rangle_\text{tree}$ vanishes.

At the level of free theory, the variations with respect to $iA_{at}$ and $A_{rt}$ correspond to the insertions of $n_0 -i\omega\chi_0\delta\varphi_r + \chi_0 A_{rt}$ and $\omega\chi_0 + i\chi_0 A_{at}$ respectively. We may diagrammatically represent these as
\begin{align}\label{eq:gcan-lin-source}
    \tikzsetnextfilename{int-Ear}
    \twoint{aux}{ext2}
    = \omega\chi_0, \qquad 
    \tikzsetnextfilename{int-aEr}
    \twoint{}{ext}
    = -i\omega\chi_0, \qquad 
    \tikzsetnextfilename{int-EaEr}
    \twoint{ext}{ext2}
    = \chi_0.
\end{align}
With these ingredients, the tree-level retarded and symmetric correlation functions are represented by the Feynman diagrams
\begin{alignat}{2}\label{eq:tree-level-correlators}
    G^\rmR_{nn}(p)_{\text{tree}}
    &= 
    ~~\tikzsetnextfilename{corr-ra-tree}
    \tikz[thick]{
        \fextin{}{-16mm};
        \fline[8mm]{bold}{aux}{-16mm};
        \fextout{2}{0mm};
    }~~
    + 
    ~~\tikzsetnextfilename{corr-ra-tree-contact}
    \tikz[thick]{
        \fextin{}{0mm};
        \fextout{2}{0mm};
    }~~
    &&= \frac{\bk^2\sigma_0}{\Delta(p)} ,
    \nn\\
    G^\rmS_{nn}(p)_{\text{tree}}
    &= ~~\tikzsetnextfilename{corr-rr-tree}
    \raisebox{-1pt}{\tikz[thick]{
        \fextin{}{-16mm};
        \fline[8mm]{bold}{bold}{-16mm};
        \fextout{}{0mm};
    }}
    &&= \frac{2}{\beta}
    \frac{\bk^2\sigma_0\fD}{\Delta_\rmS(p)}.
\end{alignat}

\paragraph*{Three-point interactions:}

Moving beyond the free theory, the SK effective action gives rise to various interactions
that contribute to the correlation functions via loop effects. Expanding \cref{eq:quantum-action} to cubic order in fluctuations, we find
\begin{align}\label{eq:3pt-result}
    S_3
    &= \int \df^d t\,\df^d\bx 
    \bigg[ \frac{\chi_1}{2} \lb  B_{rt}^2B_{at} 
    + \frac{\hbar^2}{12} B_{at}^3 \rb
    - \sigma_1 B_{rt} B^i_a \dow_t B_{ri} 
    - \frac{\hbar^2\sigma_1}{8} B_a^i B_{ai} \dow_t B_{at} \nn\\
    &\hspace{8em}
    + \frac{i\sigma_1}{\beta} \Big( B_{rt} B^i_a \fD B_{ai}
    + B_{at} B^i_r\fD B_{ai}
    - B^i_r B_{ai} \fD B_{at} \Big) \bigg].
\end{align}
This yields 3 three-point interaction vertices among dynamical fields
\begin{align}
    \tikzsetnextfilename{can_int-3pt-arr}
    \threeint{aux}{}{}
    &= \frac{i}{2} \omega_2\omega_3\lb \chi_1 i\omega_1
    + \sigma_1 \bk_1^2 \rb, \nn\\ 
    \tikzsetnextfilename{can_int-3pt-aar}
    \threeint{aux}{aux}{}
    &= - \frac{\sigma_1}{2\beta} \Big( 
    i\omega_3 \bk_1\cdot \bk_2 (\fD_1+\fD_2)
    + \lb i\omega_2(\bk_1 \cdot \bk_3) 
    - i\omega_1(\bk_2 \cdot \bk_3)  \rb
    \lb \fD_1 - \fD_2\rb
    \Big), \nn\\
    \tikzsetnextfilename{can_int-3pt-aaa}
    \threeint{aux}{aux}{aux}
    &= \frac{-\hbar^2\chi_1}{24} \omega_1\omega_2\omega_3
    - \frac{i\hbar^2\sigma_1}{24} 
    \Big( 
    \bk_1\cdot\bk_2 \omega_3^2 
    + \bk_2\cdot\bk_3 \omega_1^2 
    + \bk_3\cdot\bk_1 \omega_2^2 
    \Big).
\end{align}
Here $(p_{1,2,3})=(\omega_{1,2,3},\bk_{1,2,3})$ denote the frequencies and wavevectors on different legs of interactions, labelled as%
\tikzsetnextfilename{momentum_label_convention_3}%
\raisebox{-5pt}{\tikz[thick]{
    \draw [] (-2.1mm,0)--(0,0);
    \draw [] (0,0)--(1mm,1.8mm);
    \draw [] (0,0)--(1mm,-1.8mm);
    \node at (2mm,1mm) {\tiny 1};
    \node at (-2mm,1mm) {\tiny 2};
    \node at (2mm,-1mm) {\tiny 3};
}}. 
Similarly $\fD_{1,2,3} = \hbar\beta\omega_{1,2,3}/2\,\coth(\hbar\beta\omega_{1,2,3}/2)$. Note that the ``$aaa$'' vertex is a purely quantum artefact and does not arise in the classical theory.

We also find 10 three-point interaction vertices involving background fields 
\begin{gather}
    \tikzsetnextfilename{can_int-3pt-Earr}
    \threeint{ext}{}{}\quad 
    \tikzsetnextfilename{can_int-3pt-arEr}
    \threeint{aux}{}{ext2} \quad 
    \tikzsetnextfilename{can_int-3pt-aaEr}
    \threeint{aux}{aux}{ext2} \quad 
    \tikzsetnextfilename{can_int-3pt-Eaar}
    \threeint{ext}{aux}{} \quad 
    \tikzsetnextfilename{can_int-3pt-Eaaa}
    \threeint{ext}{aux}{aux} \nn\\
    \tikzsetnextfilename{can_int-3pt-aErEr}
    \threeint{aux}{ext2}{ext2} \quad 
    \tikzsetnextfilename{can_int-3pt-EarEr}
    \threeint{ext}{}{ext2}\quad 
    \tikzsetnextfilename{can_int-3pt-EaEaa}
    \threeint{ext}{ext}{aux}\quad 
    \tikzsetnextfilename{can_int-3pt-EaErEr}
    \threeint{ext}{ext2}{ext2} \quad 
    \tikzsetnextfilename{can_int-3pt-EaEaEa}
    \threeint{ext}{ext}{ext}
\end{gather}
The respective Feynman rules are given in \cref{app:feynman-rules-source}.

\paragraph*{Four-point interactions:} 

We can similarly expand the SK effective action \eqref{eq:quantum-action} to quartic order in fluctuations to find
\begin{align}\label{eq:4pt-result}
    \cL_4
    &= \frac{\chi_2}{6} \lb B_{rt}^3 B_{at} 
    + \frac{\hbar^2}{4} B_{rt} B_{at}^3
    \rb
    - \half \sigma_2 B_{rt}^2 B^i_a \dow_t B_{ri} 
    - \frac{\hbar^2\sigma_2}{8} B_{at}^2 B^i_a \dow_t B_{ri} 
    - \frac{\hbar^2\sigma_2}{8} B_a^i B_{ai} \dow_t(B_{at}B_{rt}) \nn\\
    &\qquad 
    + \frac{i\sigma_2}{\beta} \lb
    \half B_{rt}^2 B^i_a \fD B_{ai}
    + \frac{\hbar^2}{8} B_{at}^2 B^i_a \fD B_{ai}
    + B_{at}B_{rt} B^i_r\fD B_{ai}
    - B^i_r B_{ai} \fD(B_{at}B_{rt})\rb.
\end{align}
This leads to 4 four-point interactions among dynamical fields
\begin{align}
    \tikzsetnextfilename{int-4pt-arrr}
    \fourint{aux}{}{}{}
    &= \frac{1}{6}\omega_2\omega_3\omega_4 \lb \chi_2i\omega_1
    + \sigma_2 \bk_1^2\rb, \nn\\
    \tikzsetnextfilename{int-4pt-aarr}
    \fourint{aux}{aux}{}{}
    &= -\frac{\sigma_2}{4\beta} 
    \omega_3\omega_4
    (\bk_1\cdot \bk_2)(\fD_1+\fD_2) \nn\\
    &\qquad 
    - \frac{\sigma_2}{4\beta} 
    \lb
    \omega_2 \omega_4  (\bk_1\cdot\bk_3) \Big( \fD_1 
    - \fD_{1+3} \Big) 
    + (\text{3 combinations})
    \rb, \nn\\
    \tikzsetnextfilename{int-4pt-aaar}
    \fourint{aux}{aux}{aux}{}
    &= \frac{i\hbar^2\chi_2}{24} \omega_1\omega_2\omega_3\omega_4 \nn\\
    &\qquad
    - \frac{\hbar^2\sigma_2}{24} \omega_4 
    \lb \Big( \omega_1\omega_2
    (\bk_3\cdot \bk_4)
    - \omega_3(\omega_1+\omega_2)(\bk_1\cdot\bk_2) 
    \Big)
    + \text{(2 combinations)}
    \rb \nn\\
    \tikzsetnextfilename{int-4pt-aaaa}
    \fourint{aux}{aux}{aux}{aux}
    &= -\frac{\hbar^2\sigma_2}{96\beta}
    \Big(\omega_1\omega_2 (\bk_3\cdot\bk_4)\fD_4 
    + \text{(11 combinations)}
    \Big).
\end{align}
where the legs of interactions are labelled as%
\tikzsetnextfilename{momentum_label_convention_4}%
\raisebox{-5pt}{\tikz[thick]{
    \draw [] (-1.5mm,-1.5mm)--(1.5mm,1.5mm);
    \draw [] (-1.5mm,1.5mm)--(1.5mm,-1.5mm);
    \node at (2.5mm,1mm) {\tiny 1};
    \node at (-2.5mm,1mm) {\tiny 2};
    \node at (-2.5mm,-1mm) {\tiny 3};
    \node at (2.5mm,-1mm) {\tiny 4};
}}.
The combinations suppressed above are obtained after symmetrising over the legs of the same kind. Once again, the ``$aaar$'' and ``$aaaa$'' vertices are purely quantum artefacts.

% \begin{align}
%     \cL_4
%     &= \frac{\chi_2}{6} B_{rt}^3 B_{at} 
%     - \frac{\sigma_2}{2} B_{rt}^2 
%     \nabla\varphi_a\cdot\nabla\dot\varphi_r \nn\\
%     &\qquad 
%     + \frac{i\sigma_2}{\beta} \lb
%     \half B_{rt}^2 \nabla\varphi_a\cdot\fD\nabla\varphi_a
%     + \nabla\varphi_r \cdot \lb B_{at}B_{rt} \fD \nabla\varphi_a
%     - \nabla\varphi_a \fD(B_{at}B_{rt}) \rb \rb \nn\\
%     &\qquad 
%     + \frac{\hbar^2\chi_2}{24} B_{rt} B_{at}^3
%     - \frac{\hbar^2\sigma_2}{8} B_{at}^2 
%     \nabla\varphi_a\cdot\nabla\dot\varphi_r
%     - \frac{\hbar^2\sigma_2}{8} (\nabla\varphi_a)^2 \dow_t(B_{at}B_{rt}) \nn\\
%     &\qquad 
%     + \frac{i\hbar^2\sigma_2}{8\beta}
%     B_{at}^2\nabla\varphi_a\cdot \fD\nabla\varphi_a.
% \end{align}

We also find 20 four-point interaction vertices involving background fields 
\begin{gather}
    \tikzsetnextfilename{int-4pt-Earrr}
    \fourint{ext}{}{}{} \quad 
    \tikzsetnextfilename{int-4pt-arrEr}
    \fourint{aux}{}{}{ext2} \quad 
    \tikzsetnextfilename{int-4pt-Eaarr}
    \fourint{ext}{aux}{}{}\quad 
    \tikzsetnextfilename{int-4pt-aarEr}
    \fourint{aux}{aux}{}{ext2} \quad 
    \tikzsetnextfilename{int-4pt-Eaaar}
    \fourint{ext}{aux}{aux}{} \quad 
    \tikzsetnextfilename{int-4pt-aaaEr}
    \fourint{aux}{aux}{aux}{ext2} \quad 
    \tikzsetnextfilename{int-4pt-Eaaaa}
    \fourint{ext}{aux}{aux}{aux} \nn\\
    \tikzsetnextfilename{int-4pt-EarrEr}
    \fourint{ext}{}{}{ext2} \quad 
    \tikzsetnextfilename{int-4pt-arErEr}
    \fourint{aux}{}{ext2}{ext2} \quad 
    \tikzsetnextfilename{int-4pt-EaarEr}
    \fourint{ext}{aux}{}{ext2}\quad 
    \tikzsetnextfilename{int-4pt-aaErEr}
    \fourint{aux}{aux}{ext2}{ext2} \quad 
    \tikzsetnextfilename{int-4pt-EaEaar}
    \fourint{ext}{ext}{aux}{}\quad 
    \tikzsetnextfilename{int-4pt-EaaaEr}
    \fourint{ext}{aux}{aux}{ext2}\quad 
    \tikzsetnextfilename{int-4pt-EaEaaa}
    \fourint{ext}{ext}{aux}{aux} \nn\\
    \tikzsetnextfilename{int-4pt-EarErEr}
    \fourint{ext}{}{ext2}{ext2} \quad 
    \tikzsetnextfilename{int-4pt-aErErEr}
    \fourint{aux}{ext2}{ext2}{ext2} \quad
    \tikzsetnextfilename{int-4pt-EaEaEar}
    \fourint{ext}{ext}{ext}{} \quad 
    \tikzsetnextfilename{int-4pt-EaEaaEr}
    \fourint{ext}{ext}{aux}{ext2}  \quad
    \tikzsetnextfilename{int-4pt-EaErErEr}
    \fourint{ext}{ext2}{ext2}{ext2} \quad
    \tikzsetnextfilename{int-4pt-EaEaEaEr}
    \fourint{ext}{ext}{ext}{ext2}
\end{gather}
The respective Feynman rules are given in \cref{app:feynman-rules-source}.

\paragraph*{GGL regularisation:}

Note that the interaction vertices outlined above carry factors of frequencies. Consequently, the frequency integrals inside loops typically diverge. As such, this is typical feature of interacting field theories and warrants a careful regularisation prescription to tame such integrals. However, the issue is more subtle in SK-EFTs because improper regularisation may lead to misleading unphysical results like unstable poles in the retarded correlation functions and nonzero expectation values of noise correlation functions~\cite{Crossley:2015evo} (see also~\cite{Haehl:2016pec, Jensen:2017kzi}). 

The authors in~\cite{Gao:2018bxz} proposed a specific regularisation scheme to circumvent these issues, which we will call GGL (Gao-Glorioso-Liu) regularisation. Therein, the denominator $\Delta(p)$ in the ``$ra$'' propagator in \cref{eq:free-props} is replaced by
\begin{align}\label{eq:GGL-F}
    \frac{1}{\Delta(p)} 
    \to \frac{1}{\Delta_{\GGL}(p)}
    = \frac{1}{\Delta(p) (1-i\omega/\Lambda_\omega)^N},
\end{align}
for some positive integer $N$ and a large non-negative frequency cutoff $\Lambda_\omega$. All the frequency loop integrals can be rendered finite by choosing a large enough $N$. We can remove the cutoff at the end of our calculations by sending $\Lambda_\omega\to\infty$.
To be consistent with the KMS symmetry, we need to also simultaneously regularise the ``$rr$'' propagator.
We find that the respective denominator $\Delta_\rmS(p)$ in \cref{eq:free-props} get regularised as
\begin{align}
    \frac{\bk^2\sigma_0}{\Delta_\rmS(p)}
    \to \frac{\bk^2\sigma_0}{\Delta_{\rmS,\GGL}(p)}
    = \frac{\bk^2\sigma_0 \Re(1-i\omega/\Lambda_\omega)^p
    + \omega\chi \Im(1-i\omega/\Lambda_\omega)^p}
    {\big|\Delta(p)(1-i\omega/\Lambda_\omega)^p\big|^{2}}.
\end{align}
Note that the GGL regularisation preserves the following splitting property of the $\Delta_{\rmS}(p)$, i.e.
\begin{align}
    \frac{2\bk^2D}{\Delta_{\rmS,\GGL}(p)}
    = \frac{1}{\Delta_{\text{GGL}}(p)}
    + \frac{1}{\Delta_{\text{GGL}}(-p)},
    \label{eq:GGL-identity}
\end{align}
which we will exploit later.

% To wit, the GGL-regularised ``$ra$'' propagator is given as
% \begin{align}
%     \ftikz{prop-ra}
%     ~~
%     &= \frac{1}{\chi\omega}
%     \frac{1}{\Delta_{\text{GGL}}(p)}
%     \equiv \frac{1}{\chi\omega}
%     \frac{1}{\Delta(p)(1-i\omega/\Lambda_\omega)^p}, 
% \end{align}

\subsection{One-loop retarded correlation functions}
\label{sec:one-loop}

Let us use the ingredients outlined above to obtain one-loop corrections to the two-point retarded correlation function of density. The respective tree-level result is given in \cref{eq:tree-level-correlators}. There are two kinds of diagrams that we need to consider: diagrams involving two three-point interactions and diagrams involving one four-point interaction.

\paragraph*{Corrections from three-point interactions:}

Let us first look at one-loop diagrams involving two three-point interactions. There are 4 such diagrams involving only dynamical field interactions and 8 more involving background field interactions. They can be computed as follows
\begin{align}\label{eq:3pt-diagrams}
&\tikzsetnextfilename{corr3x3-rarr}
\retdiagramA{}{aux}{}{aux}{}{}{}{aux}
+ 
\tikzsetnextfilename{corr3x3E-rarr}
\retdiagramAp{}{aux}{}{aux}{}{}
+ 
\tikzsetnextfilename{corr3x3E2-rarr}
\retdiagramApp{}{aux}{}{}{}{aux}
+
\tikzsetnextfilename{corr3x3E3-rarr}
\retdiagramAppp{}{aux}{}{} \nn\\
&\quad 
= 
-\frac{\bk^2 \lb \chi_1 D - \sigma_1 \rb}{2\beta\chi^2\Delta(p)^2} 
\int_{p'}
\lB
\Big(
{i\omega} \bk\cdot \bk' \sigma_1
+ D\bk^2
\lb i\chi_1 \omega' + \sigma_1 \bk'\cdot \bk'' \rb \Big)
\frac{2D\bk''^2\fD''}{\Delta(p')\Delta_\rmS(p'')}
+ (p'\leftrightarrow p'')
\rB \nn\\
&\quad 
= 
-\frac{\bk^2 \lb \chi_1 D - \sigma_1 \rb}{2\beta\chi^2\Delta(p)^2} 
\int_{p'}
\lB
\Big(
{i\omega} \bk\cdot \bk' \sigma_1
+ D\bk^2
\lb i\chi_1 \omega' + \sigma_1 \bk'\cdot \bk'' \rb \Big)
\frac{\fD''}{\Delta(p')\Delta(p'')}
+ (p'\leftrightarrow p'')
\rB \nn\\
&\qquad\qquad\qquad 
\text{(GGL regularisation)}, \nn\\[0.5em]
&\tikzsetnextfilename{corr3x3-raar}
\retdiagramA{}{aux}{}{aux}{aux}{}{}{aux}
+ 
\tikzsetnextfilename{corr3x3E-raar}
\retdiagramAp{}{aux}{}{aux}{aux}{}
+ 
\tikzsetnextfilename{corr3x3E2-raar}
\retdiagramApp{}{aux}{aux}{}{}{aux}
+
\tikzsetnextfilename{corr3x3E3-raar}
\retdiagramAppp{}{aux}{aux}{} \nn\\
&\quad 
= -\frac{\bk^2\lb \chi_1 D - \sigma_1 \rb \sigma_1}{2\beta\chi^2\Delta(p)^2}
\int_{p'}\lB
\Big( D\bk^2 \bk'\cdot \bk'' \fD'
+ i\omega''\bk \cdot \bk' \lb \fD' - \fD''\rb
\Big) \frac{1}{\Delta(p')\Delta(p'')} + (p'\leftrightarrow p'')
\rB, \nn\\[.5em]
&
\tikzsetnextfilename{corr3x3-rara}
\retdiagramA{}{aux}{}{aux}{}{aux}{}{aux}
+
\tikzsetnextfilename{corr3x3E-rara}
\retdiagramAp{}{aux}{}{aux}{}{aux} 
= 0\qquad \text{(GGL regularisation)}, \nn\\[0.5em]
&
\tikzsetnextfilename{corr3x3-rara-nonunitary}
\retdiagramA{}{}{}{aux}{}{aux}{}{aux}
+
\tikzsetnextfilename{corr3x3E-rara-nonunitary}
\retdiagramAp{}{}{}{aux}{}{aux}
= 0\qquad \text{(GGL regularisation)}.
\end{align}
Here $p'=(\omega',\bk')$ and $p''=(\omega'',\bk'')$ denote the frequencies and wavevectors on the upper and lower legs of the loop respectively, with $p'' = p-p'$. Furthermore $\fD'$ and $\fD''$ denote the operator $\fD$ evaluated on the frequencies $\omega'$ and $\omega''$ respectively. We have used the shorthand notation 
\begin{align}
    \int_{p'} = \int \frac{\df\omega'\df^d\bk'}{(2\pi)^{d+1}}.
\end{align}

Note that the diagrams in the last two lines involve a purely retarded loop. Therefore they vanish under GGL regularisation for sufficiently large $N$. One the other hand, the factor of $1/\Delta_\rmS(p'')$ in the diagrams in the first line splits into $1/\Delta(p'')$ and $1/\Delta(-p'')$ parts using the identity in \cref{eq:GGL-identity}. The latter part also gives rise to a purely retarded loop and drops under GGL regularisation for sufficiently large $N$.

\paragraph*{Corrections from four-point interactions:}

Next, let us first look at one-loop diagrams involving one four-point interaction. There are 3 such diagrams involving only dynamical field interactions and 5 more involving background field interactions. Interestingly, all of these vanish under GGL regularisation
\begin{align}
&
\tikzsetnextfilename{corr4-rr}
\retdiagramC{}{aux}{}{}{}{aux}
+
\tikzsetnextfilename{corr4E-rr}
\retdiagramCp{}{aux}{}{}
+
\tikzsetnextfilename{corr4E2-rr}
\retdiagramCpp{}{}{}{aux}
+
\tikzsetnextfilename{corr4E3-rr}
\retdiagramCppp{}{} \nn\\
&\quad 
= \frac{\bk^2}{\beta\chi\Delta(p)^2}
\int_{p'}
\Big( D^2\bk^2 \chi_2 - i\omega\sigma_2 \Big)
\frac{D \bk'^2\fD'}{\Delta_\rmS(p')} \nn\\
&\quad 
= 0\qquad \text{(GGL regularisation)},
\nn\\[.5em]
&
\tikzsetnextfilename{corr4-ra}
\retdiagramC{}{aux}{}{aux}{}{aux}
+
\tikzsetnextfilename{corr4E-ra}
\retdiagramCp{}{aux}{}{aux} 
= 0\qquad \text{(GGL regularisation)}, \nn\\[.5em]
&
\tikzsetnextfilename{corr4-ra-nonunitary}
\retdiagramC{}{}{}{aux}{}{aux}
+
\tikzsetnextfilename{corr4E-ra-nonunitary}
\retdiagramCp{}{}{}{aux} 
= 0\qquad \text{(GGL regularisation)}.
\end{align}
The diagrams in the last two lines vanish for sufficiently large $N$ due to the purely retarded loop. Whereas, the factors of $1/\Delta_\rmS(p')$ in the diagrams in the first line splits into $1/\Delta(p')$ and $1/\Delta(-p')$, leading to purely retarded and purely advanced loop integrals that also vanish for sufficiently large $N$.

\paragraph*{Frequency integral:}

The total one-loop contribution to the two-point retarded correlation function of density under GGL regularisation is given as
\begin{align}
    G^\rmR_{nn}(p)
    &=
    \frac{\bk^2\sigma_0}{\Delta(p)} 
    + \frac{\bk^4\chi_0^2\lambda^2}{2\beta\Delta(p)^2} 
    \int_{p'}
    \frac{i\omega' \fD'' + i\omega''  \fD'}{\Delta(p')\Delta(p'')}
    + \ldots,
    \qquad 
    \lambda = \frac{\dow D}{\dow n} = \frac{\sigma_1 - \chi_1 D}{\chi^2_0}.
\end{align}
where ellipses denote higher-loop corrections. The frequency loop integral appearing above can be computed using residue theorem. We have
\begin{align}
    I 
    &= \int \frac{\df\omega'}{2\pi}
    \frac{i\omega' \fD'' + i\omega''  \fD'}{\Delta(p')\Delta(p'')} \nn\\
    &= \frac{\hbar\beta}{2}
    \int \frac{\df\omega'}{2\pi}
    \frac{i\omega'(\omega-\omega')}
    {\big(D\bk'^2 - i\omega'\big)
    \big(D\bk''^2 - i\omega + i\omega' \big)}
    \lB \coth\lb\frac{\hbar\beta}{2}\,(\omega-\omega')\rb
    - \coth\lb\frac{\hbar\beta}{2}\,\omega'\rb\rB.
\end{align}
One can check that the integrand falls off sufficiently fast at infinity for the residue theorem to apply, so we can skip the GGL regularisation for this integral. We can close the contour in the lower-half complex-$\omega'$ plane, where we pick up the poles at $\omega' = -iD\bk'^2$ from the denominator and at $\omega' = -2in\pi/(\hbar\beta)$ and $\omega' = \omega -2in\pi/(\hbar\beta)$ for $n\geq 0$ from the factors of $\coth$. We find
\begin{align}
    I &= 
    \frac{\hbar\beta}{2}
    \frac{D\bk'^2(D\bk'^2-i\omega)}{
    (D\bk'^2 + D\bk''^2 - i\omega)} 
    \lB \cot\lb \frac{\hbar\beta}{2}(D\bk'^2-i\omega) \rb
    - \cot\lb \frac{\hbar\beta}{2} D\bk'^2\rb \rB
    \nn\\
    &\qquad 
    - \sum_{n=1}^\infty \lb
    \frac{\omega_n(i\omega+\omega_n)}{
    (D\bk''^2 + \omega_n)(D\bk'^2 - i\omega - \omega_n)} 
    +
    \frac{\omega_n(i\omega-\omega_n)}{
    (D\bk''^2 - i\omega + \omega_n)(D\bk'^2 - \omega_n)} \rb.
\end{align}
where $\omega_n = 2\pi n/(\hbar\beta)$. The summations above are finite and can be performed in terms of Digamma functions $\psi$, i.e.
\begin{align}
    I &= 
    \frac{\hbar\beta}{2}
    \frac{D\bk'^2(D\bk'^2-i\omega)}{
    (D\bk'^2 + D\bk''^2 - i\omega)} 
    \lB \cot\lb \frac{\hbar\beta}{2}(D\bk'^2-i\omega) \rb
    - \cot\lb \frac{\hbar\beta}{2} D\bk'^2\rb \rB
    \nn\\
    &\qquad 
    + \frac{\hbar\beta}{2\pi}
    \frac{D\bk'^2(D\bk'^2-i\omega)}{D\bk'^2 + D\bk''^2 -i\omega}
    \lB \psi\lb -\frac{\hbar\beta}{2\pi} D\bk'^2\rb
    - \psi\lb -\frac{\hbar\beta}{2\pi} (D\bk'^2-i\omega)\rb \rB \nn\\
    &\qquad 
    + \frac{\hbar\beta}{2\pi}
    \frac{D\bk''^2(D\bk''^2-i\omega)}{D\bk'^2 + D\bk''^2 -i\omega}
    \lB \psi\lb \frac{\hbar\beta}{2\pi} D\bk''^2\rb
    - \psi\lb \frac{\hbar\beta}{2\pi} (D\bk''^2-i\omega)\rb \rB.
\end{align}
Using the identity $\psi(-x)=\psi(x) + \pi\cot(\pi x) + 1/x$, we can express the answer as
\begin{align}
    I &= 
    \frac{\hbar\beta}{2\pi}
    \frac{D\bk'^2(D\bk'^2-i\omega)}{D\bk'^2 + D\bk''^2 -i\omega}
    \lB \psi\lb \frac{\hbar\beta}{2\pi} D\bk'^2\rb
    - \psi\lb \frac{\hbar\beta}{2\pi} (D\bk'^2-i\omega)\rb \rB \nn\\
    &\qquad 
    + \frac{\hbar\beta}{2\pi}
    \frac{D\bk''^2(D\bk''^2-i\omega)}{D\bk'^2 + D\bk''^2 -i\omega}
    \lB \psi\lb \frac{\hbar\beta}{2\pi} D\bk''^2\rb
    - \psi\lb \frac{\hbar\beta}{2\pi} (D\bk''^2-i\omega)\rb \rB \nn\\
    &\qquad 
    - \frac{i\omega}{D\bk'^2 + D\bk''^2 -i\omega}.
\end{align}
Note that this expression is symmetric under $\bk'\leftrightarrow\bk''$ as we expect.

\paragraph*{Wavevector integral:}

To compute the wavevector integral, let us perform a redefinition of the integration variables
\begin{align}
    \bq = \bk' - \half \bk = \half\bk - \bk''.
\end{align}
so that the denominator becomes $2D(\bq^2 + \cM/4)$, where $\cM = k^2 - 2i\omega/D$. We also use the Taylor expansion of the Digammma function
\begin{align}
    \psi(z_1) - \psi(z_2)
    = \frac{z_1-z_2}{z_1z_2}
    + \sum_{n=1}^{\infty} (-1)^{\,n+1}\,\zeta(n+1)
    \Big( z_1^{n} - z_2^{n} \Big),
\end{align}
where $\zeta(n+1)$ denotes the Riemann Zeta function. Plugging these above, the integrand becomes
\begin{align}
    I &= \frac{i\omega}{2D}
    \lb \frac{1}{\bq^2 + \cM/4}
    + \sum_{n=1}^{\infty}\zeta(n+1)
    \bfrac{-\hbar\beta D}{2\pi}^{n+1} 
    \frac{\cA^{(n)}(\omega,\bq+\bk/2) 
    + \cA^{(n)}(\omega,\bq-\bk/2)}{\bq^2+\cM/4}
    \rb,
    \label{eq:I-expression}
\end{align}
where the quantum corrections are given by
\begin{align}
    \cA^{(n)}(\omega,\bq)
    &= \frac{D}{i\omega}
    \bq^2\Big(\bq^2-i\omega/D\Big)
    \lB
    \bq^{2n}
    - \Big(\bq^2-i\omega/D\Big)^n \rB \nn\\
    &= 
    \Big(\bq^2-i\omega/D\Big)
    \sum_{m=0}^{n-1} \binom{n}{m} \bq^{2m+2}
    (-i\omega/D)^{n-m-1} \nn\\
    &= 
    \sum_{m=0}^{n} c_{n,m} (-i\omega/D)^{n-m}\bq^{2m+2}.
\end{align}
We have identified the coefficients 
\begin{align}
    c_{n,m} = 
    \begin{cases}
        1 \qquad &\text{for}~m=0,  \\
        m \qquad &\text{for}~m=n,  \\
        \binom{n}{m} + \binom{n}{m-1}
        \qquad &\text{for}~0<m<n.
    \end{cases}
\end{align}

\begin{table}[t]
    \setlength{\tabcolsep}{.5em}
    \renewcommand{\arraystretch}{1.3}
    \centering
    \begin{tabular}{c|ccccc}
        \hline
         $\cJ_{d,m}(x)$ & $m{\,=\,}0$ & $m{\,=\,}1$ & $m{\,=\,}2$
         & $m{\,=\,}3$ & $m{\,=\,}4$ \\
         \hline
         $d=1$ & 1 & $1-x$ & $1-6x+x^2$ & $1-15x+15x^2-x^3$
         & $1-28x+70x^2-28x^3+x^4$ \\
         $d=2$ & 1 & $1-x$ & $1-4x+x^2$ & $1-9x+9x^2-x^3$
         & $1-16x+36x^2-16x^3+x^4$ \\
         $d=3$ & 1 & $1-x$ & $1- \frac{10}{3}x + x^2$ & $1-7x+7x^2-x^3$ 
         & $1-12x+\frac{126}{5}x^2-12x^3+x^4$ \\
         $d=4$ & 1 & $1-x$ & $1- 3x + x^2$ & $1-6x+6x^2-x^3$ 
         & $1-10x+20x^2-10x^3+x^4$ \\
         \hline
    \end{tabular}
    \caption{The $\cJ_{d,m}(x)$ polynomials for small values of $d$ and $m$.}
    \label{tab:cJ-functions}
\end{table}

\begin{table}[t]
    \setlength{\tabcolsep}{.5em}
    \renewcommand{\arraystretch}{1.3}
    \centering
    \begin{tabular}{c|ccccc}
        \hline
         $\cP_{d,m}(x)$ 
         & $m{\,=\,}1$ & $m{\,=\,}2$ & $m{\,=\,}3$ & $m{\,=\,}4$ & $m{\,=\,}5$ \\
         \hline
         $d=1$ & 0 & $1+2x+x^2$ & 0
         & $1-12x-26x^2-12x^3+x^4$ & 0 \\
         $d=2$ & 0 & $1+x^2$ & 0
         & $1-8x-4x^2-8x^3+x^4$ & 0
          \\
         $d=3$ & 0 & $1-\frac23 x+x^2$ & 0
         & $1-\frac{20}{3}x + \frac{26}{15}x^2-\frac{20}{3}x^3+x^4$ & 0 \\ 
         $d=3$ & 0 & $1-x+x^2$ & 0
         & $1-6x +4x^2-6x^3+x^4$ & 0 \\ 
         \hline
    \end{tabular}
    \caption{The $\cP_{d,m}(x)$ polynomials for small values of $d$ and $m$. Note that $\cP_{d,m}(x)$ is zero for odd $m$.}
    \label{tab:cB-functions}
\end{table}

We can perform the integration of $I$ in terms of the following standard integrals
\begin{align}
    \cJ_{d,m}(\bk^2) =
    \frac{(4\pi)^{d/2}\Gamma\big(\tfrac d2\big)}
    {\Gamma\bigl(1-\tfrac d2-m\bigr)
    \Gamma\big(\tfrac d2+m\big)}
    \int \frac{\df^d\bq}{(2\pi)^d}
    \frac{(\bq\pm\bk)^{2m}}{\bq^2+1},
\end{align}
where $\Gamma$ denotes the Gamma function.
These integrals have been normalised such that $\cJ_{d,m}(0) = 1$. We have explicitly evaluated the first few of these in \cref{tab:cJ-functions}.
% \begin{align}
%     \cJ_0(\bk^2) = 1, \qquad 
%     \cJ_1(\bk^2) = 1-\bk^2, \qquad 
%     \cJ_2(\bk^2) = 1 - \frac{10}{3}\bk^2 + \bk^4
% \end{align}
Using this, we find
\begin{align}
    \int \frac{\df^d\bq}{(2\pi)^d}
    \frac{(\bq\pm\bk/2)^{2m+2}}{\bq^2+\cM/4}
    = \bfrac{\cM}{4}^{d/2+m}
    \frac{\Gamma\bigl(-\tfrac d2-m\bigr)
    \Gamma\big(\tfrac d2+m+1\big)}{(4\pi)^{d/2}\Gamma\big(\tfrac d2\big)}
    \cJ_{m+1}(z),
\end{align}
where $z = \bk^2/\cM$.
% 
% Note that 
% \begin{align}
%     \int \frac{\df^d\bq}{(2\pi)^d}
%     \frac{(\bq\pm\bk/2)^{2m+2}}{\bq^2+\cM/4}
%     &=
%     \frac{(\cM/4)^{\,\tfrac{d}{2}+m}}{(4\pi)^{d/2}}
%     \frac{\Gamma\bigl(-\tfrac d2-m\bigr)
%     \Gamma\big(\tfrac d2+m+1\big)}{\Gamma\big(\tfrac d2\big)} \nn\\
%     &\hspace{4em}\times
%     {}_1F_1\!\Bigl(
%      -m-1,
%      -\tfrac{d}{2}-m,
%      -\tfrac{\bk^2}{\cM}
%     \Bigr),
% \end{align}
% \begin{align}
%     \int \frac{\df^d\bq}{(2\pi)^d}
%     \frac{(\bq\pm\bk/2)^{2m+2}}{\bq^2+\cM/4}
%     &=
%     \frac{(\cM/4)^{\,\tfrac{d}{2}+m}}{(4\pi)^{d/2}}
%     \frac{\Gamma\bigl(-\tfrac d2-m\bigr)
%     \Gamma\big(\tfrac d2+m+1\big)}{\Gamma\big(\tfrac d2\big)} \nn\\
%     &\hspace{4em}\times
%     {}_1F_1\!\Bigl(
%      -m-1,
%      -\tfrac{d}{2}-m,
%      -\tfrac{\bk^2}{\cM}
%     \Bigr),
% \end{align}
% To evaluate the wavevector integral, we can use the following dimensional-regularised result
% \begin{align}
%     \int \frac{\df^d\bq}{(2\pi)^d}
%     \frac{(\bq\pm\bk/2)^{2m+2}}{\bq^2+\cM/4}
%     &=
%     \int \frac{\df^d\bq}{(2\pi)^d}
%     \frac{1}{\bq^2+\cM/4}
%     \sum_{r=0,e}^{m+1}
%     \binom{m+1}{r}
%     (\bq^2 + \bk^2/4)^{m+1-r}
%     (\bq\cdot\bk)^r
% \end{align}
The $\Gamma$ factors in the expression above can be simplified using the identity $\Gamma(z)\Gamma(1-z) = \pi/\sin(\pi z)$, leading to
\begin{align}
    \frac{\Gamma\bigl(-\tfrac d2-m\bigr)
    \Gamma\big(\tfrac d2+m+1\big)}{\Gamma\big(1-\tfrac d2\big)\Gamma\big(\tfrac d2\big)}
    = \frac{\sin(\pi\tfrac d2)}{\sin\lb\pi\lb\tfrac d2+m+1\rb\rb}
    = (-1)^{m+1}.
\end{align}
For even $d$, this expression has to be understood with appropriate regularisation $d\to d+\varepsilon$. Plugging these results into the calculation of $I$ in \cref{eq:I-expression}, we find the final result
\begin{align}
    \int \frac{\df^d\bq}{(2\pi)^d} I 
    &= \frac{2i\omega}{D}
    \frac{\cM^{\,\tfrac{d}{2}-1}}{(16\pi)^{d/2}}
    \Gamma\bigl(1-\tfrac d2\bigr)
    \lb 1
    - 2\sum_{n=1}^{\infty}\zeta(n+1)
    \bfrac{\hbar\beta\cM}{8\pi D}^{n+1} 
    \cP_{d,n+1}(z)
    \rb,
\end{align}
where the polynomials $\cP_{d,n+1}$ are defined as
\begin{align}
    \cP_{d,n+1}(z)
    &= \frac{(-1)^{n}(4\pi)^{d/2}}{\Gamma\bigl(1-\tfrac d2\bigr)}
    \bfrac{\cM}{4}^{-{\tfrac d2}-n}
    \int \frac{\df^d\bq}{(2\pi)^d}
    \frac{\cA^{(n)}(\omega,\bq\pm\bk/2)}{\bq^2+\cM/4} \nn\\
    &= 
    - \sum_{m=0}^{n} c_{n,m}
    2^{n-m} \lb z - 1\rb^{n-m}\cJ_{m+1}(z) \nn\\
    &= 
    -2^n(z-1)^{n+1}
    \sum_{m=1}^{n} {\binom{n}{m-1}}
    \lB \frac{1}{2^{m-1}}
    \frac{\cJ_{m}(z)}{\lb z - 1\rb^{m}}
    + \frac{1}{2^m}
    \frac{\cJ_{m+1}(z)}{\lb z - 1\rb^{m+1}}\rB.
\end{align}
We have evaluated the first few of these polynomials in \cref{tab:cB-functions}.
We find that the polynomials $\cP_{d,n+1}$ are identically zero for even $n$. Therefore, the final result for one-loop corrections to the two-point retarded correlation function of density is given in \cref{eq:long-time-tails-quantum}.

\section{Outlook}
\label{sec:outlook}

In this work, we have derived a quantum-complete SK-EFT for the hydrodynamics of a single scalar charge, consistent with quantum FDT requirements, by enforcing the KMS symmetry at nonzero $\hbar$. In the classical limit, the KMS symmetry only enforces fluctuations at quadratic order in the noise fields in the SK effective action. However, upon including quantum corrections, we find that the KMS symmetry generates fluctuation terms at all orders in the noise fields. We have used our theory to calculate one-loop quantum corrections to the two-point retarded correlation function of the charge density at arbitrary order in $\hbar$. In doing so, we have evaluated quantum corrections to hydrodynamic long-time tails in the diffusion model at the one-loop level. While our general result is expressed as a sum over a set of polynomials, we have also been able to find a closed-form expression for the quantum corrections at small wavevector.

The polynomials $\cP_{d,2n}(z)$ in our results are defined through specific wavevector integrals. However, at the time of writing, we have not been able to identify these polynomials in terms of standard families of polynomials in the literature. It would be quite interesting to make progress on this front. Such an identification may also help us to resum the results in \cref{eq:long-time-tails-quantum} for finite wavevector and thus probe the full analytic structure of the quantum corrections. Furthermore, it will be useful to extend these computations to higher-loop orders and to include higher-derivative corrections in the diffusion equation.

We have focused on a simple hydrodynamic model of a diffusive scalar conserved charge. An immediate generalization of this work would be to derive the quantum-complete SK effective action for full hydrodynamics, including fluid velocity and momentum fluctuations. For classical long-time tails, it is known that the non-analytic corrections arising from shear viscosity enter the retarded correlation functions earlier in the derivative expansion as compared to the diffusion model~\cite{Kovtun:2003vj}. In view of this, it will be interesting to explore the analytic structure of quantum corrections in full hydrodynamics including momentum modes, based on the works of~\cite{Grozdanov:2013dba, Harder:2015nxa, Crossley:2015evo, Haehl:2015uoc, Haehl:2018lcu, Jensen:2017kzi}, as well as extensions to non-relativistic hydrodynamics~\cite{Jain:2020vgc, Armas:2020mpr}. In this regard, it will also be interesting to investigate the stability and causality issues of relativistic hydrodynamics in the quantum-complete SK-EFT paradigm~\cite{Jain:2023obu}.

We mentioned in the main text that the SK effective action we have derived is likely not unique as a quantum completion of the diffusion model. More generally, SK-EFTs may admit non-universal fluctuation terms that are not constrained by the KMS symmetry~\cite{Jain:2020zhu}. It will be of interest to explore how this non-universality behaves in the presence of quantum fluctuations. Regardless, it is known that FDTs, or equivalently the KMS symmetry, are sufficiently constraining to entirely fix the SK-EFT effective action up to cubic interactions. Since our result for quantum corrections to the retarded correlation function and quantum long-time tails is only sensitive to cubic couplings, our conclusions are therefore robust under such non-universalities.

As mentioned in the introduction, quantum fluctuations are particularly important in theories or models that lack any intrinsic scale of their own, such as conformal field theories. In this context, holography, or the AdS/CFT correspondence~\cite{Maldacena:1997re}, offers an interesting paradigm for testing the predictions of our model. In particular, it may be possible to derive the quantum-complete version of the SK effective action using holography, following the work of~\cite{deBoer:2018qqm, Glorioso:2018mmw}.

As a final remark, we note that we have used the GGL prescription to regulate frequency loop integrals in the SK-EFT~\cite{Gao:2018bxz}. A more systematic approach to regularization in SK-EFTs involves introducing additional ghost fields to counter the divergences. This approach also enables us to read off interesting cutoff-dependent results needed to study non-equilibrium RG flows, which are identically set to zero under GGL regularization. In the classical limit, the ghost part of the SK effective action is determined by an emergent BRST supersymmetry~\cite{Gao:2017bqf, Jensen:2018hse}. However, it is not clear at the moment whether such a symmetry principle extends to the quantum level. As a future direction, it will be interesting to identify the ghost contributions to the quantum-complete SK effective action, which may be used to regulate frequency integrals instead of the GGL prescription. We leave these investigations for future work.

\acknowledgements

The author would like to thank Pavel Kovtun, Jay Armas, Blaise Gout\'eraux, Shiraz Minwalla, and Loganayagam R for various useful discussions.
The author is supported by the ERC Consolidator Grant
GeoChaos (101169611).
This work is funded by the European Union.
Views and opinions expressed are however those of the
authors only and do not necessarily reflect those of the
European Union or the European Research Council Executive Agency. Neither the European Union nor the
granting authority can be held responsible for them. For
the purpose of open access, the authors have applied a
CC BY public copyright licence to any Author Accepted
Manuscript (AAM) version arising from this submission.

\makereferences

\appendix 

\section{Feynman rules for background field couplings}
\label{app:feynman-rules-source}

In this appendix we outline the Feynman rules for three-point and four-point couplings involving background fields in the quantum-complete SK-EFT for diffusion.
The Feynman rules for three-point background field couplings arising from \cref{eq:3pt-result} are given as
\begin{align}
    \ftikz[-18]{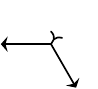}
    &= -\frac{\chi_1}{2}\omega_2\omega_3, \qquad 
    \ftikz[-18]{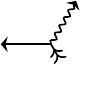}
    = - i\chi_1 \omega_1\omega_2
    + \sigma_1 \omega_2 (\bk_1\cdot \bk_2), \nn\\
    \ftikz[-18]{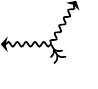}
    &= \frac{\sigma_1}{2\beta} (\bk_1 \cdot \bk_2) (\fD_1+\fD_2), \qquad 
    \ftikz[-18]{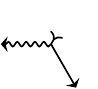}
    = \frac{i\sigma_1}{\beta} (\bk_2\cdot \bk_3) \lb \fD_1 - \fD_2 \rb, \nn\\
    \ftikz[-18]{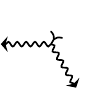}
    &= -\frac{\hbar^2\chi_1}{8}\omega_2\omega_3 
    + \frac{i\hbar^2\sigma_1}{8} (\omega_2+\omega_3) \bk_2\cdot \bk_3, \nn\\
    \ftikz[-18]{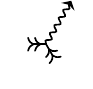}
    &= \frac{\chi_1}{2}\omega_1, \qquad
    \ftikz[-18]{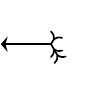}
    = -i\chi_1\omega_2, \qquad 
    \ftikz[-18]{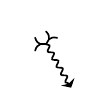}
    = - \frac{\hbar^2\chi_1}{8}\omega_3, \nn\\
    \ftikz[-18]{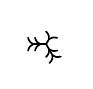}
    &= \frac{\chi_1}{2}, \qquad
    \ftikz[-18]{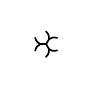}
    = -\frac{\hbar^2\chi_1}{24}.
\end{align}
On the other hand, the Feynman rules for four-point background field couplings arising from \cref{eq:4pt-result} are given as
\begin{align}
    \ftikz[-18]{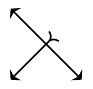}
    &= \frac{i\chi_2}{6}\omega_2\omega_3\omega_4, \qquad 
    \ftikz[-18]{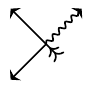}
    = -\frac{\chi_2}{2}\omega_1\omega_2\omega_3
    - \frac{i\sigma_2}{2} \omega_2\omega_3 
    \Big( \bk_1\cdot\bk_2 + \bk_1\cdot\bk_3 \Big)
    , \nn\\
    \ftikz[-18]{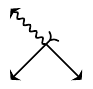}
    &= - \frac{\sigma_2}{2\beta}
    \Big(
    \omega_4 (\bk_2\cdot\bk_3) (\fD_2 - \fD_{2+3})
    + \omega_3 (\bk_2\cdot\bk_4) (\fD_2 - \fD_{2+4})
    \Big)
    , \nn\\
    \ftikz[-18]{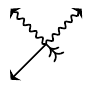}
    &= - \frac{i\sigma_2}{2\beta} \omega_3 (\bk_1\cdot\bk_2)
    (\fD_1+\fD_2) \nn\\
    &\hspace{2em}
    - \frac{i\sigma_2}{2\beta} 
    \Big(\omega_2 (\bk_1\cdot\bk_3) (\fD_1-\fD_{1+3})
    + \omega_1 (\bk_2\cdot\bk_3) (\fD_2-\fD_{2+3})
    \Big)
    ,\nn\\
    \ftikz[-18]{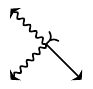}
    &= \frac{i\hbar^2\chi_2}{8}\omega_2\omega_3\omega_4
    - \frac{\hbar^2\sigma_2}{8} \omega_4
    \Big(\omega_2 (\bk_3\cdot\bk_4)
    + \omega_3 (\bk_3\cdot\bk_4)
    \Big)
    + \frac{\hbar^2\sigma_2}{8}
    \omega_4 (\omega_2+\omega_3)
    (\bk_2\cdot\bk_3), \nn\\
    \ftikz[-18]{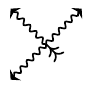}
    &= - \frac{\hbar^2\chi_2}{24}\omega_1\omega_2\omega_3
    + \frac{i\hbar^2\sigma_2}{24} 
    \Big( \omega_3 (\omega_1+\omega_2) (\bk_1\cdot\bk_2) + \text{(2 combinations)}
    \Big)
    , \nn\\
    \ftikz[-18]{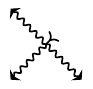}
    &= - \frac{\hbar^2\sigma_2}{48\beta} \omega_2 \Big( (\bk_3\cdot\bk_4)\fD_4
    + \text{(5 combinations)}
    \Big) , \nn\\
    \ftikz[-18]{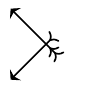}
    &= -\frac{\chi_2}{2}\omega_2\omega_3, \qquad 
    \ftikz[-18]{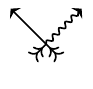}
    = - \frac{i\chi_2}{2}\omega_1\omega_2
    + \frac{\sigma_2}{2} \omega_2 (\bk_1\cdot\bk_2), \nn\\
    \ftikz[-18]{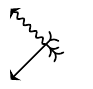}
    &= - \frac{i\sigma_2}{\beta}\omega_1 (\bk_2\cdot\bk_3)
    (\fD_2 - \fD_{2+3})
    , \qquad 
    \ftikz[-18]{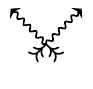}
    = \frac{\sigma_2}{4\beta} (\bk_1\cdot\bk_2)
    (\fD_1+\fD_2),  \nn\\
    \ftikz[-18]{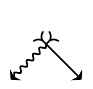}
    &= \frac{i\hbar^2\chi_2}{8} \omega_3\omega_4
    - \frac{\hbar^2\sigma_2}{8}\omega_4 (\bk_3\cdot\bk_4)
    , \nn\\
    \ftikz[-18]{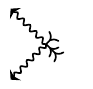}
    &= - \frac{\hbar^2\chi_2}{8} \omega_2\omega_3
    + \frac{i\hbar^2\sigma_2}{8} (\omega_2+\omega_3)
    (\bk_2\cdot\bk_3)
    , \quad
    \ftikz[-18]{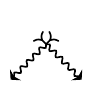}
    = - \frac{\hbar^2\sigma_2}{16\beta} (\bk_3\cdot\bk_4)
    (\fD_3+\fD_4), \nn\\
    \ftikz[-18]{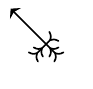}
    &= - \frac{i\chi_2}{2}\omega_2, \quad 
    \ftikz[-18]{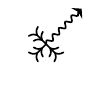}
    = \frac{\chi_2}{6}\omega_1, \quad
    \ftikz[-18]{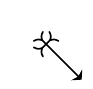}
    = \frac{i\hbar^2\chi_2}{24}\omega_4, \quad 
    \ftikz[-18]{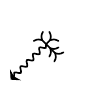}
    = - \frac{\hbar^2\chi_2}{8}\omega_3,  \nn\\
    \ftikz[-18]{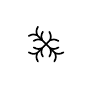}
    &= \frac{\chi_2}{6}, \qquad
    \ftikz[-18]{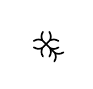}
    = -\frac{\hbar^2\chi_2}{24}.
\end{align}

\end{document}